\documentclass[usenatbib,useAMS,letterpaper]{mn2e}
\usepackage{graphicx}
\usepackage[fleqn]{amsmath}
\usepackage{amssymb}
\usepackage[totalwidth=480pt, totalheight=680pt]{geometry}

\newcommand\apj{ApJ}
\newcommand\apjl{ApJL}
\newcommand\apjs{ApJS} 
\newcommand\mnras{MNRAS} 
\newcommand\aap{A\&A} 
\newcommand\aaps{A\&AS} 
\newcommand\nat{Nature}

\newcommand\icarus{Icarus}

\title[Migration in gravitoturbulent discs]{Rapid inward migration of planets formed by gravitational instability}

\author[C. Baruteau et al.]{Cl{\'e}ment Baruteau$^{1,2}$\thanks{E-mail: 
C.Baruteau@damtp.cam.ac.uk}
, Farzana Meru$^{3,4}$ and Sijme-Jan Paardekooper$^{1}$
\\
$^{1}$DAMTP, University of Cambridge, Wilberforce Road, Cambridge CB3 0WA, UK\\
$^{2}$Department of Astronomy and Astrophysics, University of California, Santa Cruz, CA 95064, USA\\
$^{3}$School of Physics, University of Exeter, Stocker Road, Exeter, EX4 4QL, UK\\
$^{4}$Institut f{\"u}r Astronomie und Astrophysik, Universit{\"a}t T{\"u}bingen, Auf der Morgenstelle 10, 72076 T{\"u}bingen, Germany
}

\begin{document}

\date{Accepted 2011 June 1.  Received 2011 May 30; in original form 2011 March 25}

\pagerange{\pageref{firstpage}--\pageref{lastpage}} \pubyear{2011}

\maketitle

\label{firstpage}

\begin{abstract}
  The observation of massive exoplanets at large separation ($\gtrsim
  10$ AU) from their host star, like in the HR 8799 system, challenges
  theories of planet formation. A possible formation mechanism
  involves the fragmentation of massive self-gravitating discs into
  clumps. While the conditions for fragmentation have been extensively
  studied, little is known of the subsequent evolution of these giant
  planet embryos, in particular their expected orbital migration.
  Assuming a single planet has formed by fragmentation, we investigate
  its interaction with the gravitoturbulent disc it is embedded in.
  Two-dimensional hydrodynamical simulations are used with a simple
  prescription for the disc cooling. A steady gravitoturbulent disc is
  first set up, after which simulations are restarted including a
  planet with a range of masses approximately equal to the clump's
  initial mass expected in fragmenting discs. Planets rapidly migrate
  inwards, despite the stochastic kicks due to the turbulent density
  fluctuations. We show that the migration timescale is essentially
  that of type I migration, with the planets having no time to open a
  gap. In discs with aspect ratio $\sim 0.1$ at their forming
  location, planets with a mass comparable to, or larger than
  Jupiter's can migrate in as short as $10^4$ years, that is, about 10
  orbits at 100 AU. Massive planets formed at large separation from
  their star by gravitational instability are thus unlikely to stay in
  place, and should rapidly migrate towards the inner parts of
  protoplanetary discs, regardless of the planet mass.
\end{abstract}


\section{Introduction}
\label{sec:intro}
The large majority of exoplanets have been discovered by the radial
velocity and transiting techniques. Recent progress in the adaptive
optics technique has revealed about 20 exoplanets by direct
imaging. Since the latter requires a high brightness contrast between
the planet and the parent star, massive planets have been observed at
separations $\gtrsim 10$ AU \citep[e.g.,][]{Kalas08, Marois08,
  Lagrange10, Lafreniere10, Marois10}.  One such example is the HR
8799 system \citep{Marois10}, which comprises four planets with masses
evaluated between 7 and 10 Jupiter-mass, and estimated separations of
14, 24, 38 and 68 AU, respectively.

How can the large orbital separations of such massive planets be
explained? With the standard core accretion formation scenario
\citep[e.g.,][]{Safronov69, Mizuno80, Pollack96, IdaLin1}, it is
difficult to form Jupiter-like planets in isolation further than $\sim
10$ AU from a Sun-like star. Still, the formation of a first
gap-opening planet close to the snow line \citep{IdaLin1} could pave
the way for the formation of a second-generation planet at the outer
edge of the first planet's gap \citep{Bryden00, Thommes05,
  IdaLin4}. Provided that the second planet opens a gap, its outer
edge could also trigger the formation of a third-generation
planet. Subsequent planet formation could proceed as a sequence, as
long as there are enough planetesimals and gas left.

The tidal interaction between a planet and its nascent protoplanetary
disc alters the planet's semi-major axis, causing its orbital
migration \citep{gt80}. Assuming the central object is a Sun-like
star, a planet typically more massive than Saturn opens a gap around
its orbit and generally migrates inwards. It is thus unlikely that a
massive planet formed through the core-accretion scenario within $\sim
10$ AU of its host star could migrate to a separation as large as 100
AU.  A notable exception was proposed by \cite{Crida09}, who showed
that the tidal interaction between a pair of resonant massive planets
embedded in a common gap, and their protoplanetary disc, could drive
both planets to orbital separations much larger than that of their
birth place, providing some specific conditions on the planets mass
ratio and on the disc's turbulent viscosity.

Another in-situ formation scenario is based on the gravitational
instability \citep[e.g.,][]{Cameron78, Boss97}. It involves
early-stage protoplanetary discs, which are massive enough for its
self-gravity to play a part in its evolution. The probability that
discs form fragments is described by two dimensionless parameters. One
is the Toomre Q-parameter, $Q = c_{\rm s} \kappa / \pi G \Sigma$,
where $G$ is the gravitational constant, $c_{\rm s}$ and $\Sigma$
denote the disc's sound speed and surface mass density, and $\kappa$
is the epicyclic frequency. For Keplerian discs, $\kappa$
approximately equals the angular frequency $\Omega$. \cite{toomre64}
showed that, for an infinitesimally thin disc to fragment, $Q \lesssim
1$. The second parameter describing the stability of a
self-gravitating disc is the disc's cooling timescale in units of the
orbital timescale, $\beta = t_{\rm cool}\,\Omega$, where $t_{\rm cool}
= e (de / dt)^{-1}$ is the cooling timescale and $e$ the disc's
thermal energy density \citep{Gammie01}. The critical value of the
cooling parameter $\beta$, below which fragmentation will occur if $Q
\lesssim 1$, depends on the disc's adiabatic index \citep{RLA05}, and
on the local disc to primary mass ratio \citep{MeruBate10b}.  Its
exact value is currently somewhat uncertain as \cite{MeruBate10c} have
recently shown that previous simulations that determined this critical
value were unresolved. However, current simulations show that the
critical value for $\beta$ could be in the range $3 - 18$
\citep{Gammie01, RLA05, MeruBate10c, MeruBate10b}.

The vast majority of studies of planet formation by gravitational
instability have focused on the conditions to fragment discs into
bound objects. It is thought to operate at separations larger than
$30-50$ AU from the central (Sun-like) star \citep[e.g.,][]{Rafikov05,
  Matzner05, Stamatellos08}. However, little is known of the evolution
of these objects, in particular their expected migration resulting
from the gravitational interaction with the gravitoturbulent disc they
are embedded in. A few numerical studies have observed that clumps
could drift inwards \citep[e.g.,][]{Mayer02, boss2005, cha10}, and
some of them have focused on the early phases of disc formation and
evolution following the collapse from the prestellar core stage
\citep{vb06, vb10a, vb10b, Machida11}.  As these giant planet embryos
migrate inwards, they may become tidally disrupted (when their radius
becomes comparable to the clump's Hill radius), thereby delivering a
variety of planets in the inner parts of protoplanetary discs
\citep{Nayak10}.
  
In this paper, we examine the orbital evolution of a single planet
embedded in its nascent gravitoturbulent disc, following the
assumption that the planet has formed by gravitational instability. We
show that the planet migrates inwards in a timescale very similar to
that of type I migration \citep{pp09a} in the absence of
turbulence. Our physical and numerical models are described in
Section~\ref{sec:model}. In Section~\ref{sec:results}, we present our
results of simulations. Conclusions and future directions are drawn in
Section~\ref{sec:conclusion}.

\section{Model}
\label{sec:model}
We investigate the tidal interaction between a massive planet and the
protoplanetary disc it formed in through the gravitational instability
scenario. Once a gaseous clump has formed by fragmentation, it
contracts until $H_2$ dissociation occurs, causing a rapid collapse of
the clump down to planetary sizes. Our study does not address the
formation of the planet, and we consider the clump and the planet as
one object. In the following, our system of interest (clump and
planet) is referred to as the planet. Our study focuses on its
migration driven by the interaction with the disc, following the
assumption that the planet has formed by gravitational instability. As
described below, our approach is to first set up a steady state
gravitoturbulent disc, then include a single planet and follow its
orbital migration.

The disc's properties after fragmentation are poorly constrained. They
depend on how many clumps form, and how massive they are. In our
study, we assume that the disc properties (density, temperature) are
virtually unchanged by the fragmentation stage. In particular, we
assume that the disc remains gravitoturbulent.  This assumption is
motivated by the findings of previous numerical studies on disc
fragmentation \citep[e.g.,][]{MeruBate10b}, where the images of the
discs show similar structures after the formation of a clump.

\begin{figure*}
  \centering
  \includegraphics[width=0.32\hsize]{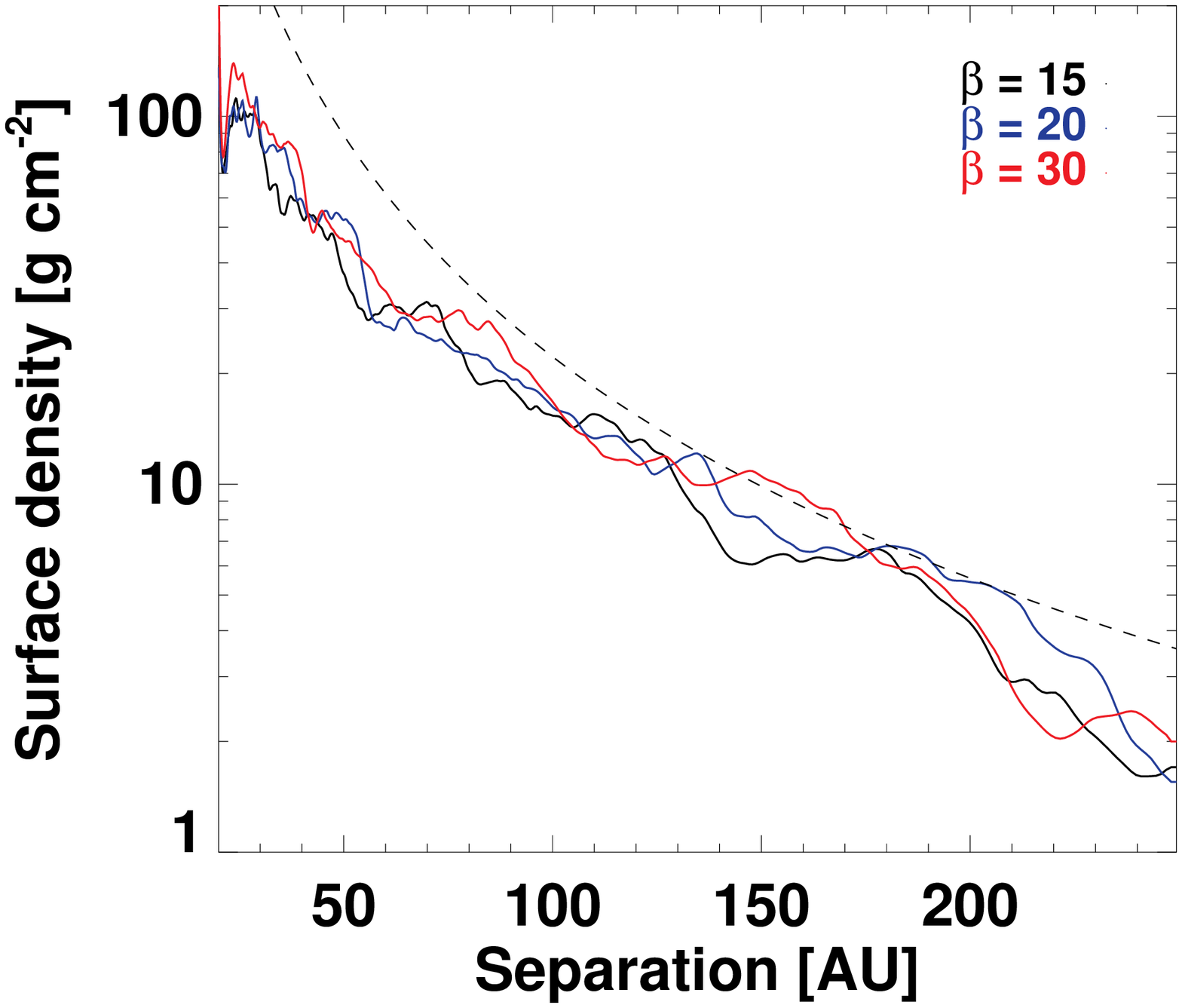}
  \includegraphics[width=0.32\hsize]{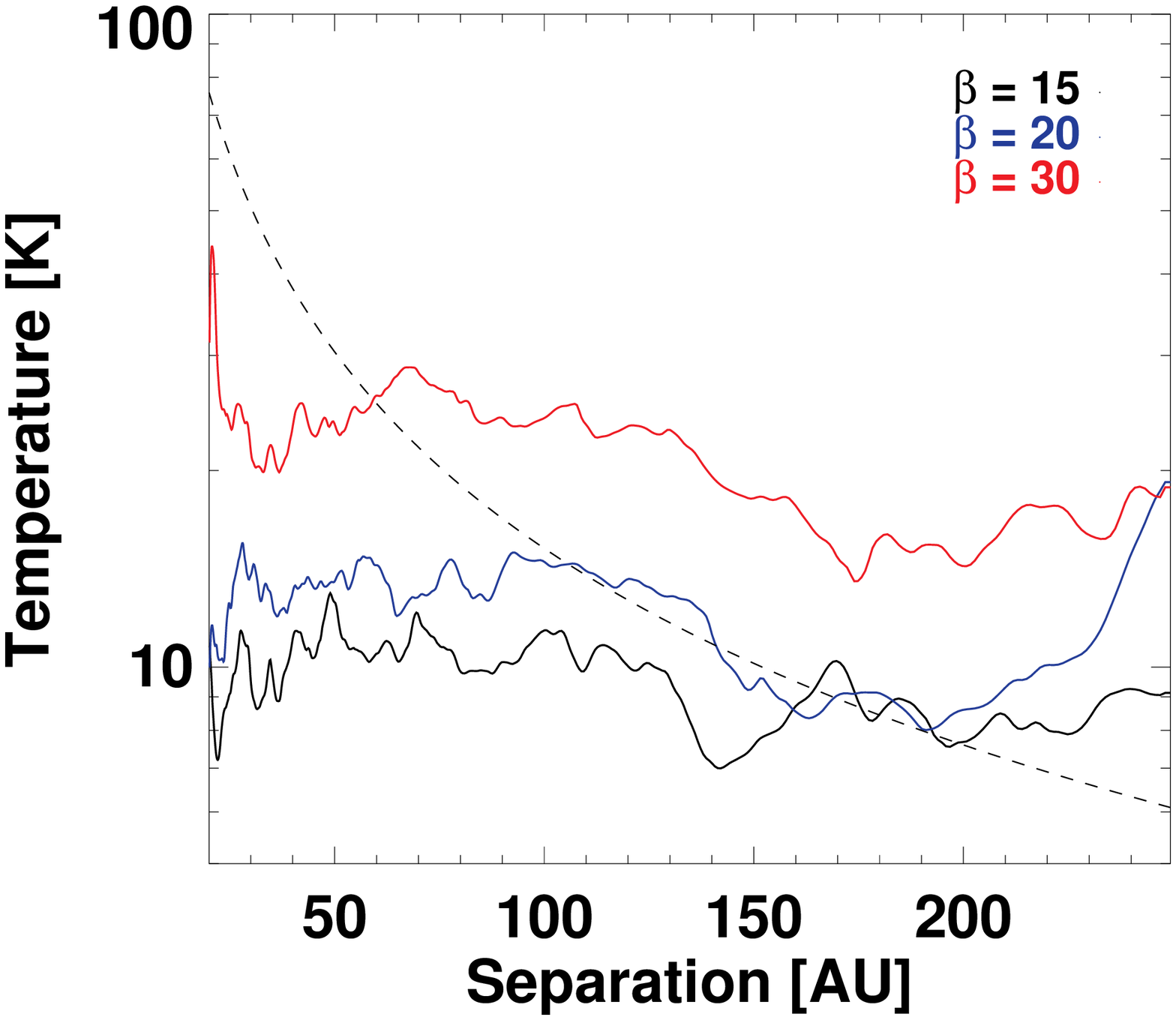}
  \includegraphics[width=0.32\hsize]{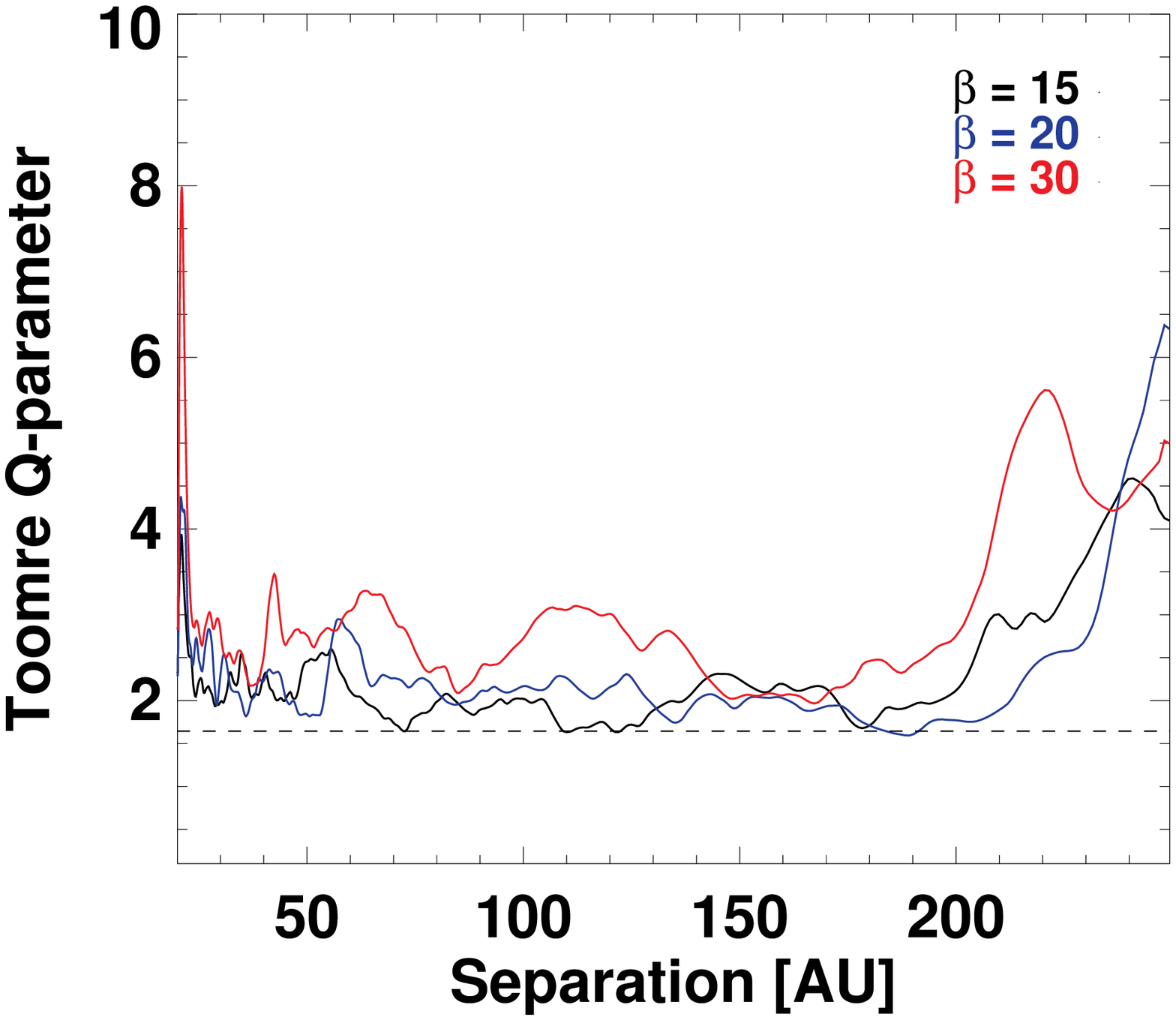}
  \caption{\label{fig:profiles}Azimuthally-averaged surface density
    (left panel), temperature (middle panel), and Toomre Q-parameter
    (right panel) obtained at 30 orbits for our three values of
    $\beta = t_{\rm cool}\,\Omega$ (solid curves). The azimuthal
    average of $Q$ was calculated as $\langle c_{\rm s}\rangle
    \langle \Omega \rangle / \pi G \langle \Sigma \rangle$, where
  $\langle . \rangle$ stands for the azimuthal average. Dashed
  curves depict the initial value of each quantity.}
\end{figure*}
We carried out two-dimensional (2D) hydrodynamical simulations using
the grid-based, staggered-mesh code FARGO\footnote{See
  \texttt{fargo.in2p3.fr}.}  \citep{fargo1}, for which a self-gravity
solver based on fast Fourier transforms was presented in
\cite{bm08b}. When calculating the self-gravitating acceleration, all
mass is assumed to be confined to the disc midplane (the razor-thin
disc approximation). The thermal energy density $e$ satisfies the
equation
  \begin{equation}
    \frac{\partial e}{\partial t} + {\bf \nabla} \cdot (e {\bf v}) 
    = -(\gamma-1) e {\bf \nabla} \cdot {\bf v} 
    + Q^{+}_{\rm bulk}
    - \frac{e}{\tau_{\rm cool}},
  \label{eqnenergy}
  \end{equation}
  where ${\bf v}$ denotes the gas velocity, and $\gamma$ is the
  adiabatic index \citep[its value can be mapped to a
  three-dimensional adiabatic index, see][and references
  therein]{Gammie01}. In Eq.~(\ref{eqnenergy}), the cooling timescale
  $\tau_{\rm cool} = \beta \Omega^{-1}$, with $\beta$ taken to be
  constant, and the heating source term $Q^{+}_{\rm bulk}$ is provided
  by artificial viscous heating at shocks arising from the
  gravitoturbulence. A von Neumann-Richtmyer artificial bulk viscosity
  is used, as described in \cite{zeus}, where the coefficient $C_2$ is
  taken equal to 1.4 ($C_2$ measures the number of zones over which a
  shock is spread over by the artificial viscosity).  Providing
  angular momentum transport driven by gravitoturbulence occurs
  locally, which is not necessarily the case \citep{bp99, Cossins09},
  the alpha parameter associated with the angular momentum flux
  density is given by \citep{Gammie01}
\begin{equation}
  \alpha = \frac{4}{9} \frac{1}{\gamma(\gamma-1)\beta},
\label{alpha_gammie}
\end{equation}
which indicates that varying $\beta$ is equivalent to varying the
viscous stress in the disc.  Preliminary calculations with our
fiducial setup (initial profiles and grid's resolution are described
below) showed that the critical value of $\beta$ below which
fragmentation occurs ranges from 10 to 15. Since we require the disc
to be in a gravitoturbulent state, but we do not require further
fragmentation so as to focus on the migration of a single planet, we
simulated three disc models with $\beta = 15$, 20 and 30.

Each of our three models initially comprises a $0.4 M_{\odot}$ 2D disc
around a star of fixed mass $M_{\star} = 1M_{\odot}$, and spanning a
radial range $20 < r < 250$ AU. (As will be shown in
Section~\ref{sec:gts}, the disc mass rapidly decreases to about
$0.2-0.25 M_{\odot}$ after 30 orbits at 100 AU). The initial surface
mass density decreases as $r^{-2}$, and it equals $\approx 20$ g
cm$^{-2}$ at 100 AU. The initial disc's temperature is $\approx 15$ K
at 100 AU, and it decreases as $r^{-1}$ (we assume a mean-molecular
weight $\mu = 2.4$ and an adiabatic index $\gamma=5/3$). This
corresponds to setting the initial disc aspect ratio $h$ (pressure
scale height to radius ratio $H/r$) equal to $0.1$ at 100 AU. The
disc's initial Toomre Q-parameter is thus uniform, equal to 2. A small
level ($\sim 0.1\%$) of white noise is added initially to break the
disc's axisymmetry. Here and in the following, disc and planet
quantities are expressed in physical units. Since our simulations are
scale-free, all the results presented throughout this paper can be
easily rescaled by using different units of mass, length and
temperature.

In all our simulations, the grid is covered with $512$ and $1536$
cells along the radial and azimuthal directions, respectively. A
logarithmic spacing is used along the radial direction, as required by
the self-gravity module in FARGO. Grid cells are approximately square,
and the initial pressure scale height is resolved by about 20 cells
along each direction. The radial and azimuthal components of the
self-gravitating acceleration are smoothed over a softening length,
$\varepsilon_{\rm sg}$. We take $\varepsilon_{\rm sg}=3\times
10^{-4}\,r$ throughout this study, which is a factor $3\times 10^{-3}$
times the initial pressure scale height.  The influence of a large
$\varepsilon_{\rm sg}$, which can be seen as a crude treatment of the
effects of finite thickness, will be examined in
Sect.~\ref{sec:numerics}, where we find that the choice for
$\varepsilon_{\rm sg}$ does not significantly alter the orbital
evolution of a planet embedded in a gravitoturbulent disc.  Outflow
boundary conditions are used at the grid's inner and outer
edges. Hydrodynamical equations are solved in the frame centred onto
the central star. Since a non-inertial frame of reference is adopted,
the indirect terms in the expression for the disc's gravitational
potential are included.

\section{Results of hydrodynamical simulations}
\label{sec:results}
Our numerical simulations were carried out in two steps. In a first
step, a gravitoturbulent disc model was set up, whose properties are
described in Section~\ref{sec:gts}. We then restarted the simulations,
including planets of several masses, and following the time-evolution
of their orbital separation. The results of the restart simulations
are presented in Section~\ref{sec:migration}. Further investigation on
the impact of the disc's temperature follows in
Section~\ref{sec:temp}.

\subsection{Gravitoturbulent stage}
\label{sec:gts}
We describe in this section the results of our three disc models with
cooling parameters $\beta=15$, 20 and 30. The absence of heating
source initially leads to decrease the disc's temperature, and thus
the Toomre Q-parameter, until gravitoturbulence sets in.  As both the
initial Q-profile and $\beta$ are taken to be uniform (the cooling
timescale thus scales proportional to the orbital timescale),
gravitoturbulence progressively develops through the disc, starting
from the inner edge. After a few orbits at the disc's outer edge, a
thermal quasi-equilibrium is reached over the whole disc, where
cooling is approximately balanced by shock heating due to
gravitoturbulence.

The disc's initial surface density that we assume does not correspond
to a mechanical equilibrium with the gravitoturbulent stress. The disc
therefore adjusts both its surface density and temperature profiles to
satisfy a mechanical equilibrium along with a uniform Toomre-Q
profile. This is illustrated in Fig.~\ref{fig:profiles}, where the
azimuthally-averaged surface mass density, temperature, and Toomre
Q-parameter are depicted at 30 orbits at 100 AU.  Their initial value
is shown with a dashed curve.  In this quasi-steady state, the surface
density profile decreases approximately as $r^{-3/2}$, and the disc's
temperature is almost uniform, as is the Toomre Q-parameter.  We
comment that the high initial disc mass ($0.4 M_{\odot}$) promotes
global modes that, along with the outflow boundary condition at the
grid's inner edge, lead to a rapid decrease in the disc mass. After 30
orbits at 100 AU, the latter ranges from $0.21M_{\odot}$ to
$0.24M_{\odot}$ for $\beta$ increasing from 15 to 30. The rate at
which the disc mass decreases is maximum at the onset of
gravitoturbulence, and then declines (but does not vanish) as a
quasi-steady state is attained.

\begin{figure}
    \includegraphics[width=\hsize]{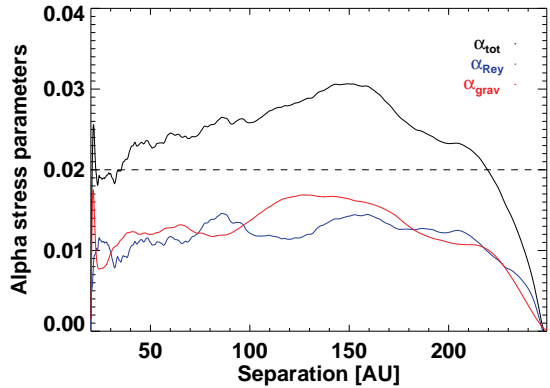}
    \caption{\label{fig:alphas}Gravitational, Reynolds and total alpha
      parameters, azimuthally- and time-averaged from 30 to 60 orbits
      at 100 AU (that is, over $\approx 8$ orbits at 250 AU, the
      location of the disc's outer edge). Results are displayed for
      the disc model with cooling parameter $\beta=20$. The dashed
      curve shows the total alpha parameter expected from viscous
      transport theory, given by Eq.~(\ref{alpha_gammie}).}
\end{figure}
The alpha parameters associated with the Reynolds and gravitational
turbulent stresses are denoted by $\alpha_{\rm Rey}$ and $\alpha_{\rm
  sg}$, respectively. Details of their calculation in a 2D disc model
are given in the Appendix. Fig.~\ref{fig:alphas} displays the
azimuthally- and time-averaged radial profiles of $\alpha_{\rm Rey}$,
$\alpha_{\rm sg}$ and of $\alpha_{\rm tot} = \alpha_{\rm sg} +
\alpha_{\rm Rey}$. Time average is done from 30 to 60 orbits at 100 AU
(that is, over $\approx 8$ orbits at the grid's outer edge). Results
are shown for $\beta=20$. We note that $\alpha_{\rm sg}$ and
$\alpha_{\rm Rey}$ take very similar values over the whole disc. The
same behavior was obtained by \cite{Gammie01} with 2D local
shearing-sheet calculations. The decrease in both $\alpha_{\rm sg}$
and $\alpha_{\rm Rey}$ at large ($\gtrsim 200$ AU) separation is most
probably due to the decreasing grid's resolution, as a logarithmic
spacing is used along the radial direction.  We comment that
$\alpha_{\rm tot}$ is locally up to $50\%$ larger than the value
expected from viscous (local) transport theory, given by
Eq.~(\ref{alpha_gammie}). The difference is probably due to our large
disc mass ($\gtrsim 0.2 M_{\star}$) and large aspect ratio ($h \sim
0.1$), which tend to foster global transport\footnote{Similar
  gravitoturbulent disc models, but with reduced aspect ratio in a
  steady state, have been simulated by \cite{pbm11} with the same
  hydrodynamical code and similar grid resolution. Using the same
  method to calculate the stresses as described in the appendix, they
  find an averaged value of $\alpha_{\rm tot}$ in very good agreement
  with the value given in Eq.~(\ref{alpha_gammie}).} of angular
momentum \citep[e.g.,][]{Lodato04}. Although not shown here, very
similar results were obtained for $\beta=15$ and $\beta=30$.

Finite resolution effects (numerical diffusion) leads to a negligible
transport of angular momentum compared to gravitoturbulence \citep[see
e.g.,][who estimated the numerical "alpha viscosity" in the FARGO code
to be $\lesssim 10^{-5}$ for a number of grid cells similar to
ours]{linpapa10}. Artificial bulk viscosity most likely leads to a
small, if not negligible transport of angular momentum compared to
gravitoturbulence, but note that it is unclear how its contribution
can be isolated in our simulations. We point out that if artificial
bulk viscosity induced a net transport of angular momentum on average,
it would lead to a discrepancy between the value of $\alpha_{\rm sg} +
\alpha_{\rm Rey}$ measured in local, steady-state disc models and the
alpha value given in Eq.~(\ref{alpha_gammie}). \cite{Gammie01} carried
out 2D local shearing-sheet simulations of gravitoturbulent disc
models using the ZEUS code, which is very similar to the FARGO code.
In particular, both codes use the same prescription for the artificial
bulk viscosity. He found that $\alpha_{\rm sg} + \alpha_{\rm Rey}$
agrees with the value given by the expression in
Eq.~(\ref{alpha_gammie}) to within $\sim 10\%$, which shows that
artificial bulk viscosity should lead to a small transport of angular
momentum compared to gravitoturbulence. Also, although not shown here,
we have checked that increasing the amount of artificial bulk
viscosity \citep[the coefficient $C_2$ in][]{zeus} by a factor of 3
leads to a negligible change in the values of $\alpha_{\rm Rey}$ and
$\alpha_{\rm sg}$.

\subsection{Restart with an embedded planet}
\label{sec:migration}
We restarted the simulations of Section~\ref{sec:gts} at 30 orbits,
including a single planet at $100$ AU. The disc's temperature and
surface mass density at this location are close to their initial value
for our three values of $\beta$ (see Fig.~\ref{fig:profiles}). Note in
particular that the disc aspect ratio at the planet's initial location
is $\approx 0.1$.  Akin to the clump's contraction timescale, the
final mass of planets formed by gravitational instability is uncertain
\citep[e.g.,][]{Boley10, Helled11}, and may substantially differ from
the clump's initial mass. The latter is estimated by \cite{Boley10} as
$\sim h_p^3\,M_{\star}$ for $Q \sim 1.5$, with $h_p$ the disc aspect
ratio at the clump's forming location. This estimate of the clump's
initial mass, which is based on the fragmentation of spiral arms, is
typically a factor $5-10$ smaller than the Toomre mass
\citep[see][their figure 1, and references therein]{Boley10}.  We
considered three planet masses corresponding to a planet to primary
mass ratio $q = M_p / M_{\star}$ around $h_p^3$: a Saturn-mass planet
($q=3\times 10^{-4}, q/h_p^3 \sim 0.3$), a Jupiter-mass planet
($q=10^{-3}, q/h_p^3 \sim 1$), and a 5 Jupiter-mass planet ($q=5\times
10^{-3}, q/h_p^3 \sim 5$). The planets gravitational potential is
softened over a smoothing length $\varepsilon = 3\times 10^{-2}\,r_p$,
with $r_p$ the planet's orbital separation, and it takes its maximum
value at the restart time (we do not gradually ramp up the planet's
mass). We will show in Section~\ref{sec:numerics} that our results are
not altered by a more gentle introduction of the planet's potential in
the disc over a dynamical timescale, the typical formation timescale
of a clump. We comment that in self-gravitating discs, the mass that
is relevant for disc--planet interactions is the mass of the planet
and of the gas envelope surrounding it. Although for simplicity the
planet's mass is fixed, gas can be accreted onto its envelope
(circumplanetary disc), thereby contributing to increasing the
effective planet's mass. Note however that this accretion rate is
probably not realistic due to our simple prescription for the disc
cooling. Since the disc is fully self-gravitating, the calculation of
the force exerted on the planet takes all the disc into account, it
does not partly exclude the planet's Hill radius \citep{cbkm09}.

\begin{figure}
  \includegraphics[width=\hsize]{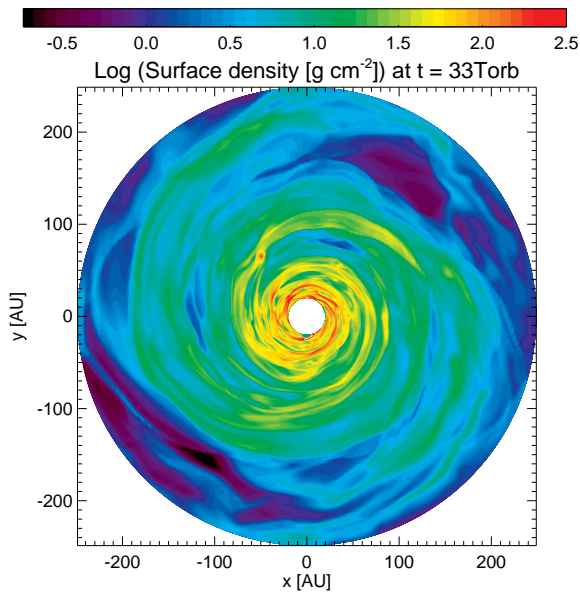}
  \caption{\label{fig:nicedens}Contours of the disc's surface density
    obtained in a restart simulation with $\beta = 30$, where a
    Jupiter-mass planet has been introduced at 100 AU. Results are
    displayed three orbits after the restart time. The planet is now
    located at about 60 AU from the central star (at $x\sim-60$ AU,
    $y\sim60$ AU).}
\end{figure}
Fig.~\ref{fig:nicedens} displays the surface density obtained for the
disc model with $\beta=30$, perturbed by a Jupiter-mass
planet. Density contours are shown only three orbits after the
restart, the planet is now located at $\approx 60$ AU from the central
object. The density perturbation due to the planet's potential is
comparable to the turbulent density perturbations. This similarity
stresses that the magnitude of the stochastic torque acting on the
planet can be similar to that of the disc--planet tidal torque
obtained in the absence of turbulence \citep[e.g.,][]{qmwmhd4, bl10}.
In such turbulent configurations, the migration timescale must be
evaluated in a statistical way, by running several simulations. For
each planet mass and each value of $\beta$, we performed a suite of
eight simulations, varying the planet's azimuth at restart as
$\varphi_p = \pi/4 \times i, \,i \in [0-7]$ (that is, independently of
the disc's density structure at restart).  The migration timescales
obtained for these series of runs are presented in
Section~\ref{sec:timescale}. The impact of stochastic density
fluctuations is described in Section~\ref{sec:kicks}. A comparison
with predictions of laminar disc models follows in
Section~\ref{sec:laminar}, and a brief discussion on the impact of
numerics is given in Section~\ref{sec:numerics}.

\subsubsection{Migration timescale}
\label{sec:timescale}
\begin{figure*}
  \centering\resizebox{\hsize}{!}
  {
    \includegraphics{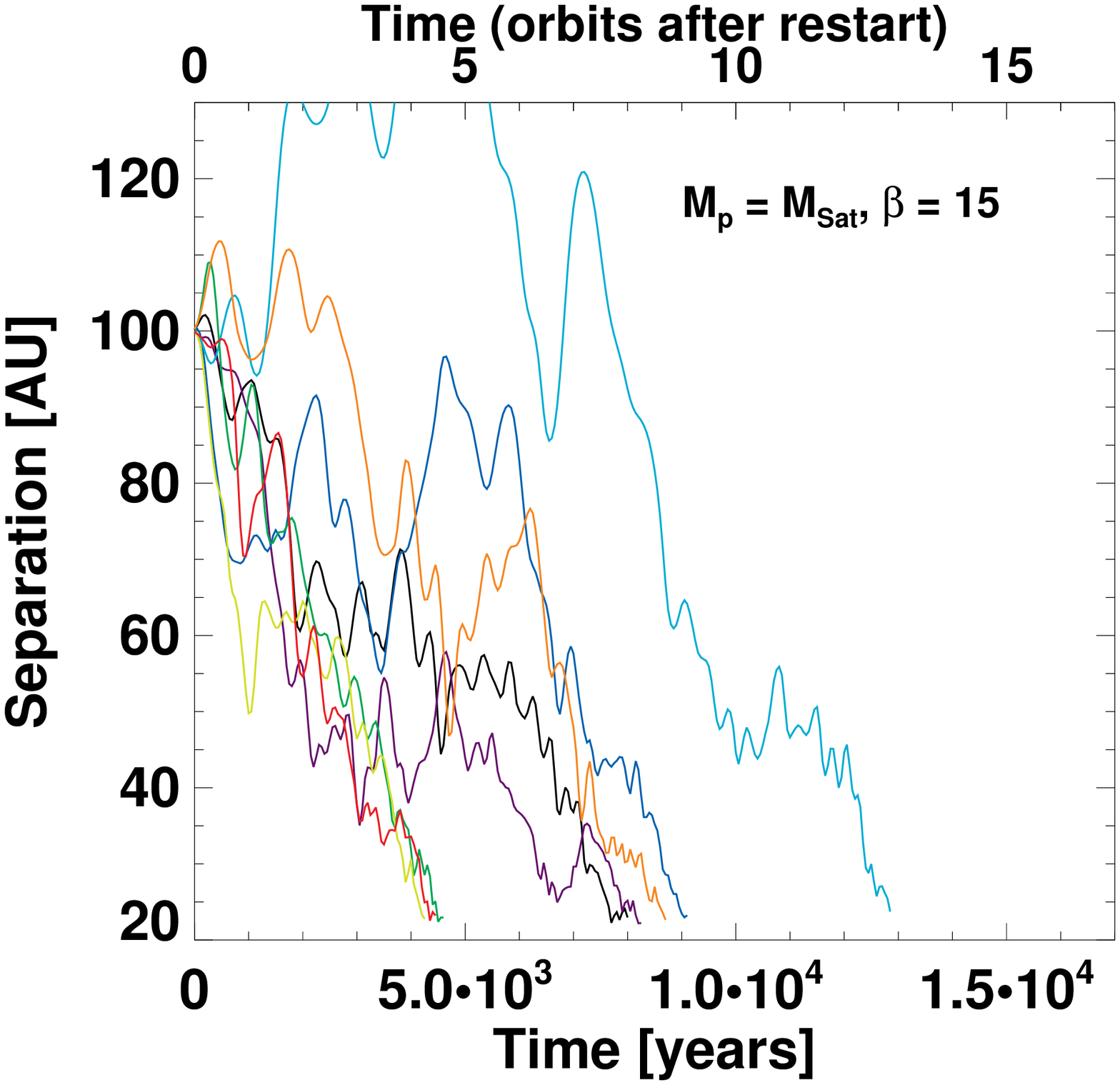}
    \includegraphics{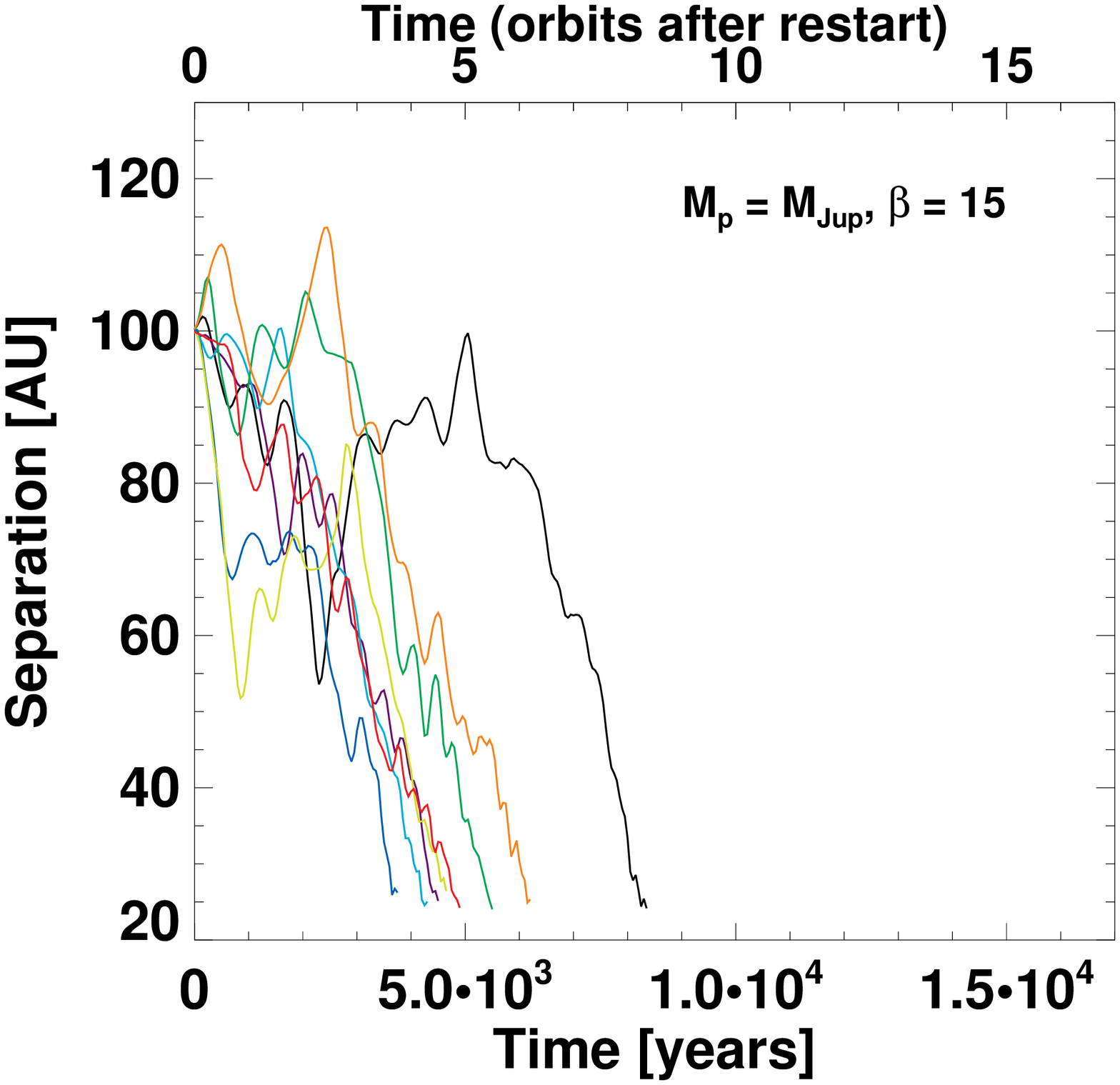}
    \includegraphics{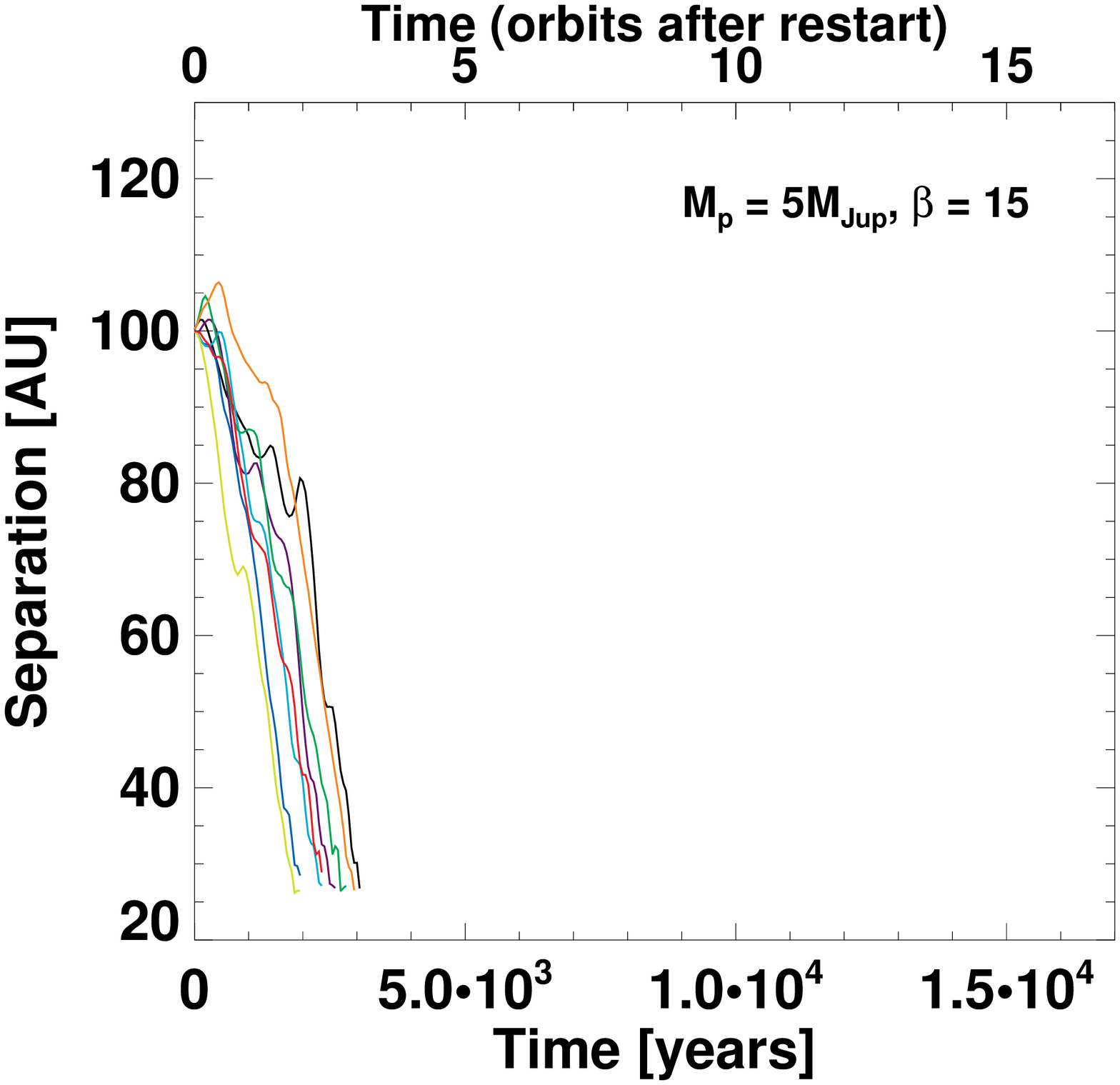}
  }
  \centering\resizebox{\hsize}{!}
  {
    \includegraphics{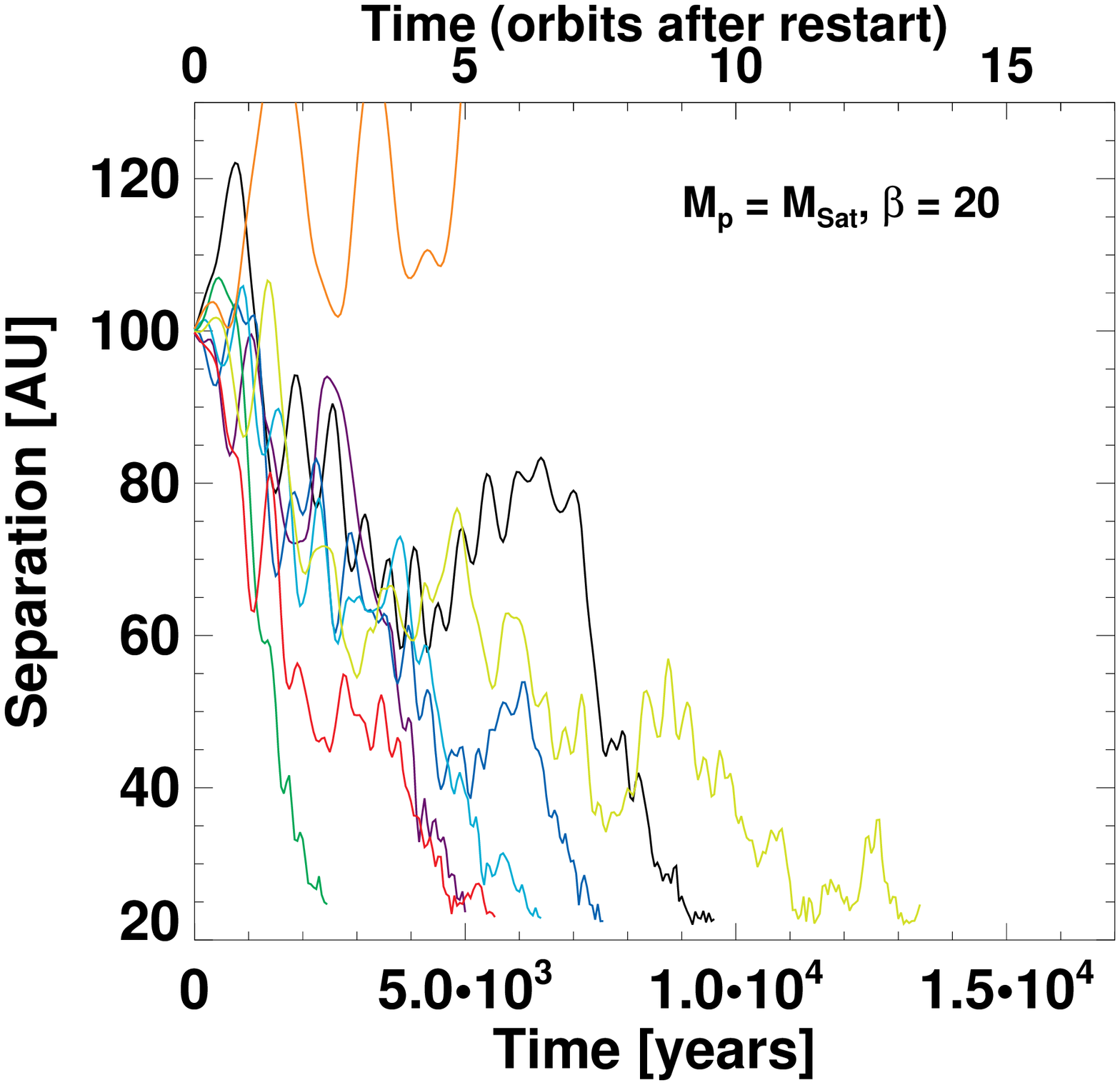}
    \includegraphics{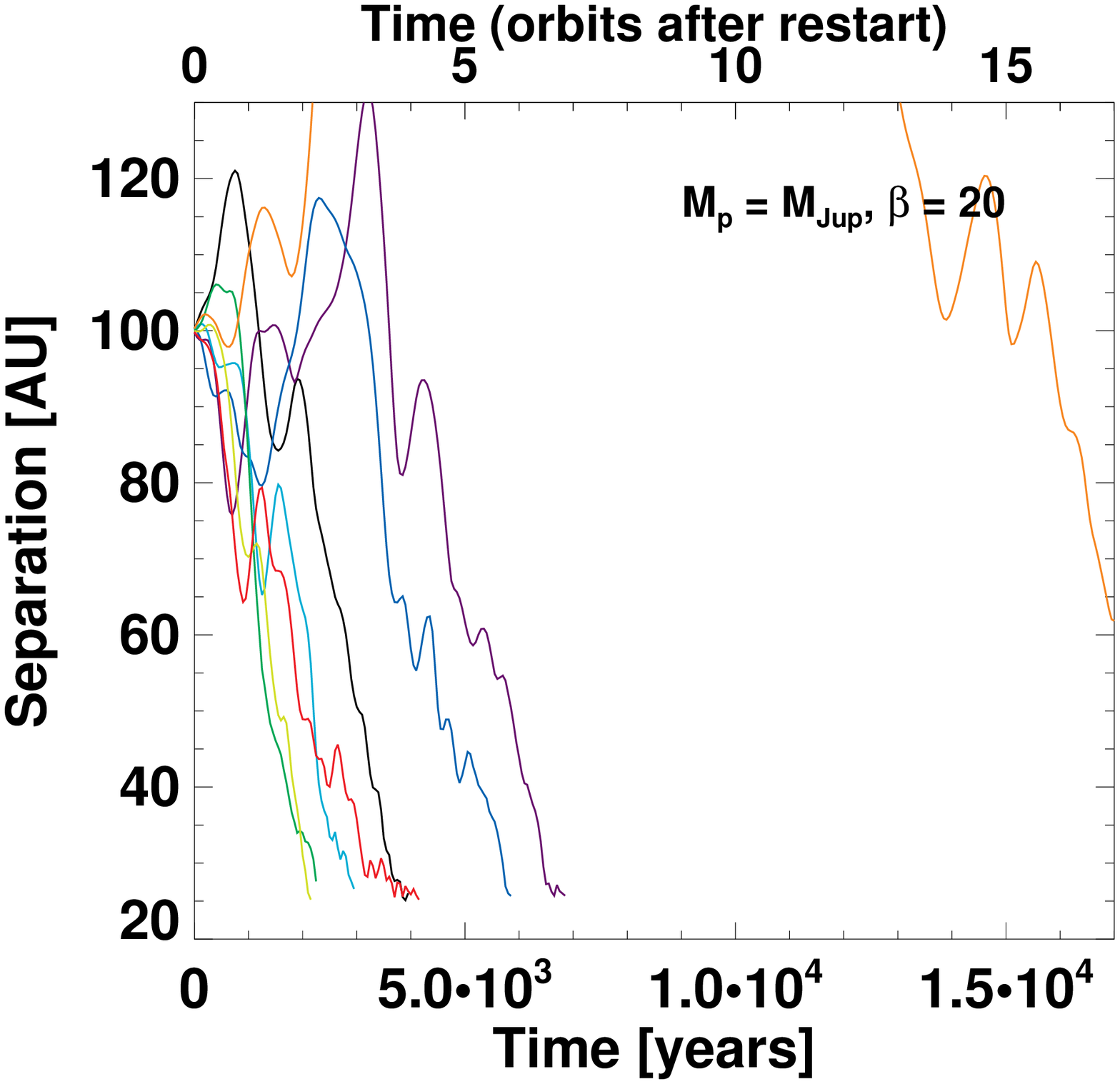}
    \includegraphics{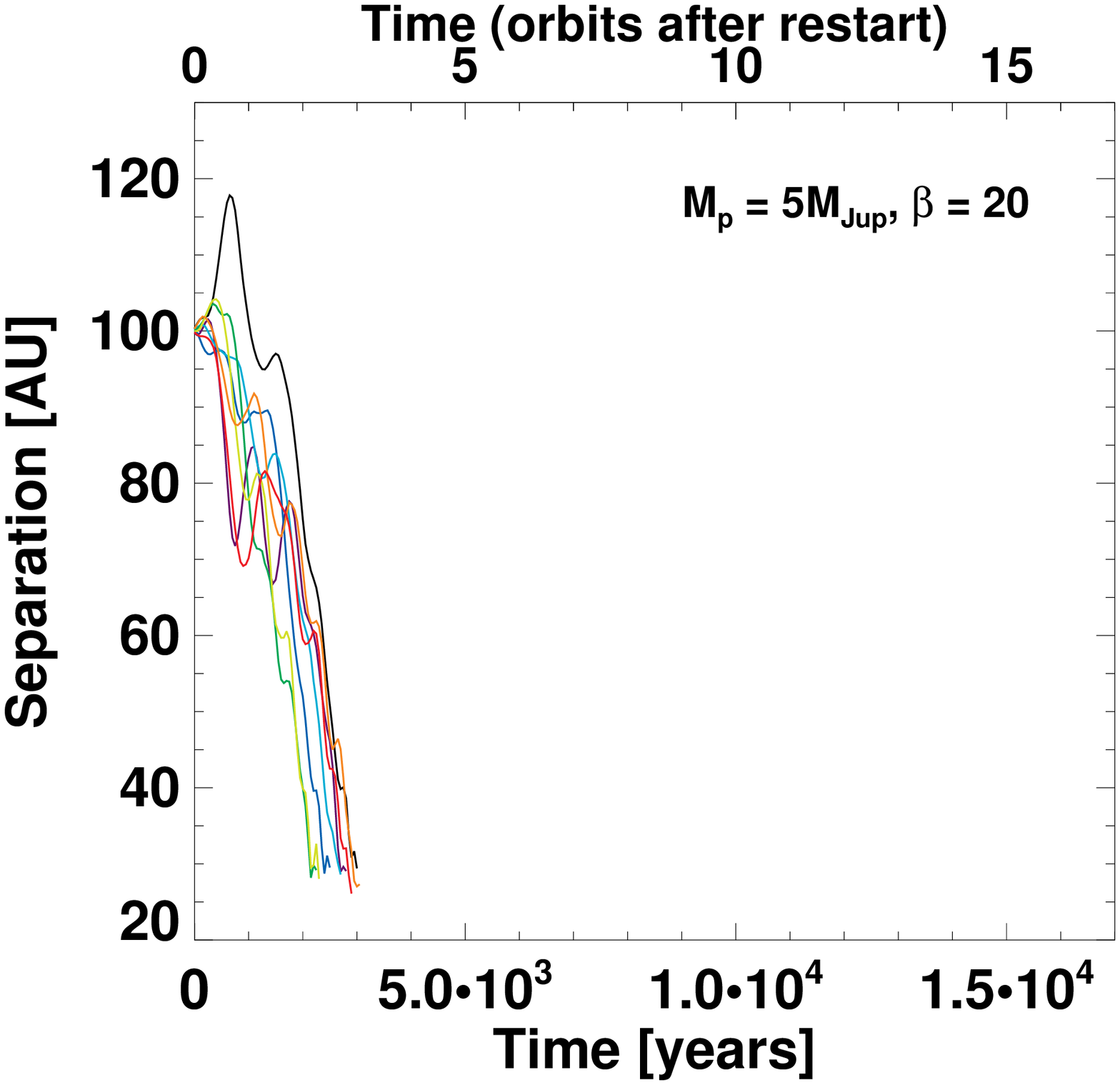}
  }
  \centering\resizebox{\hsize}{!}
  {
    \includegraphics{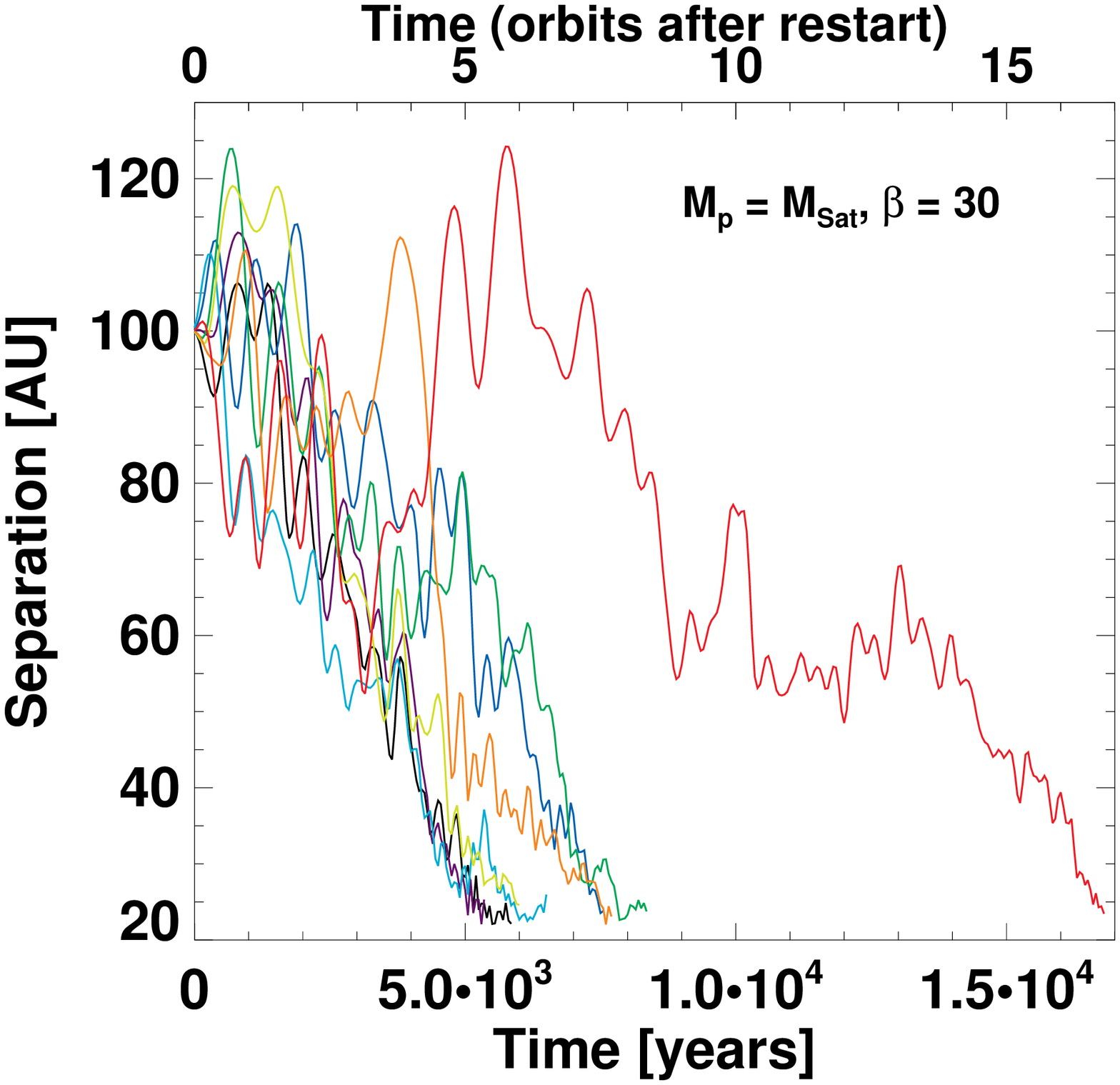}
    \includegraphics{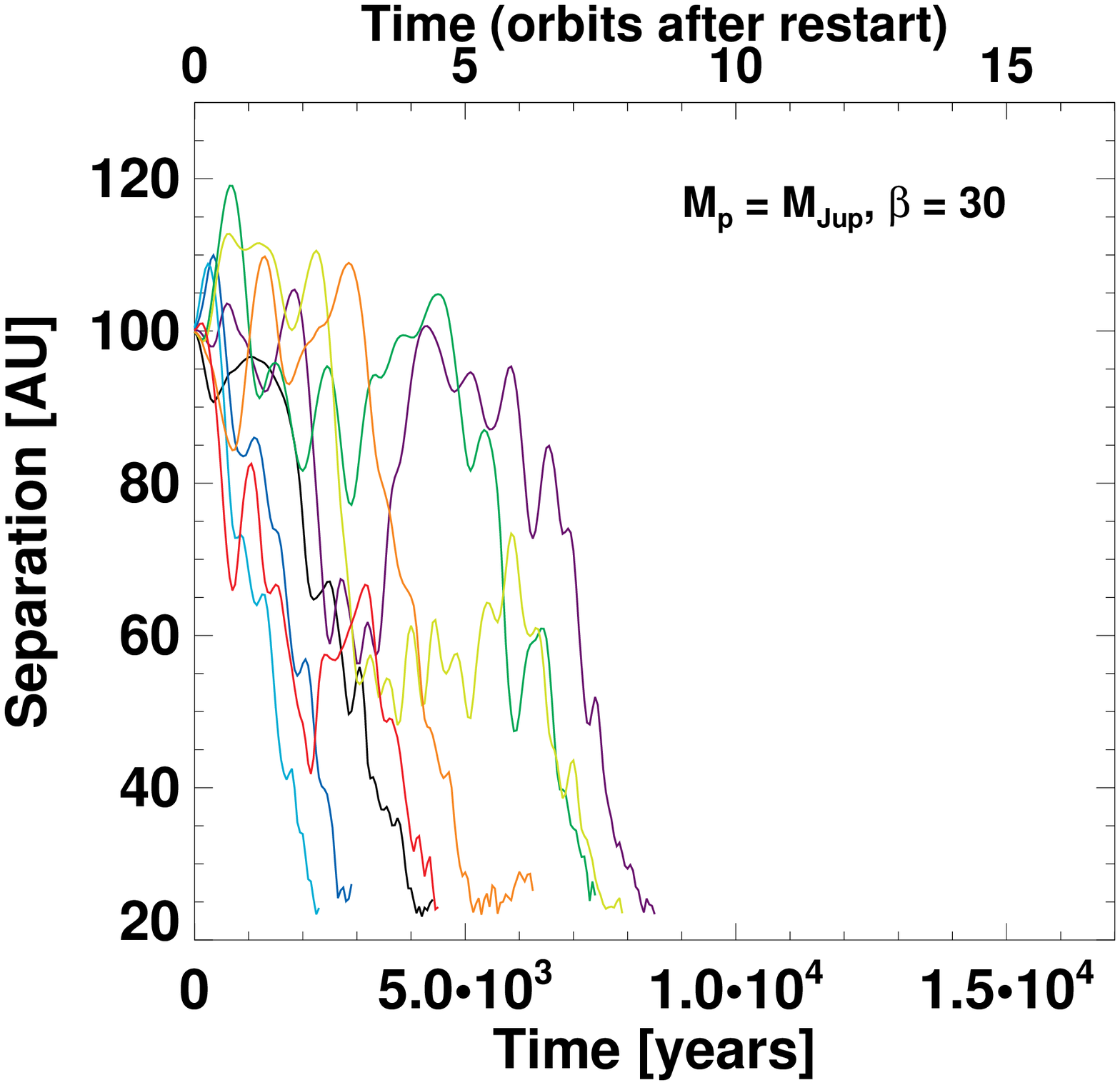}
    \includegraphics{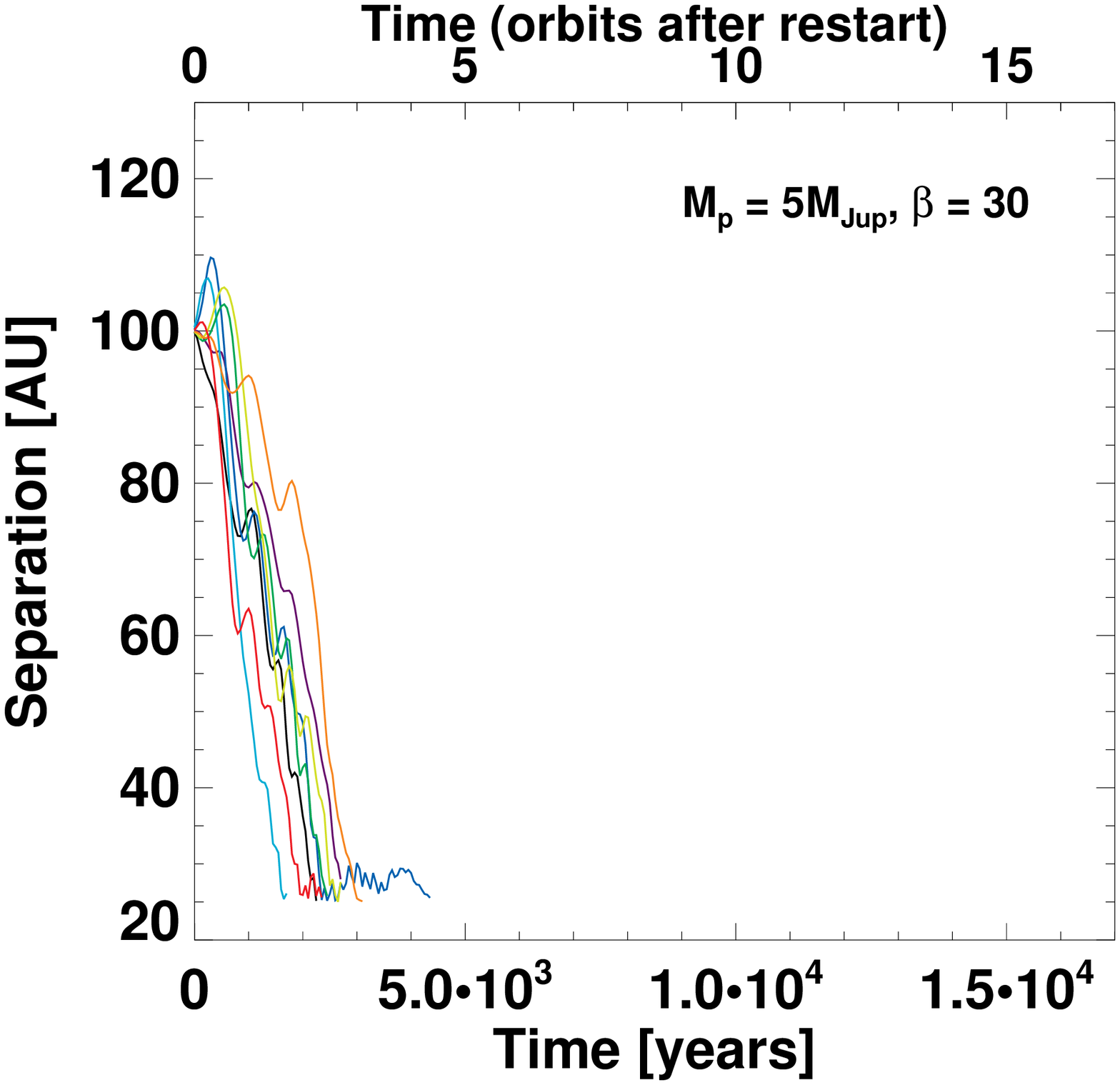}
  }
  \caption{\label{fig:azimuth}Time evolution of the orbital separation
    of a Saturn-mass planet (left column), a Jupiter-mass planet
    (middle column), and of a 5 Jupiter-mass planet (right column) in
    a gravitoturbulent disc with $\beta = 15$ (top row), $\beta = 20$
    (middle row), and $\beta=30$ (bottom row). Time is displayed in
    years (bottom x-axis), and in orbital periods at 100 AU (top
    x-axis). Results are obtained for eight evenly-spaced planet's
    azimuths at restart (azimuth increases with wavelength's color).}
\end{figure*}

The time evolution of the planets' orbital separation is displayed in
Fig.~\ref{fig:azimuth}. Planet mass increases from left to right, and
the cooling parameter $\beta$ increases from top to bottom. In all
panels, time is expressed in years (bottom x-axis), and in orbital
periods at 100 AU (top x-axis). Results are shown for the eight
evenly-spaced values of the planet's azimuth $\varphi_p$ at restart
(increases with increasing color's wavelength).
\begin{figure*}
  \centering
  \includegraphics[width=0.32\hsize]{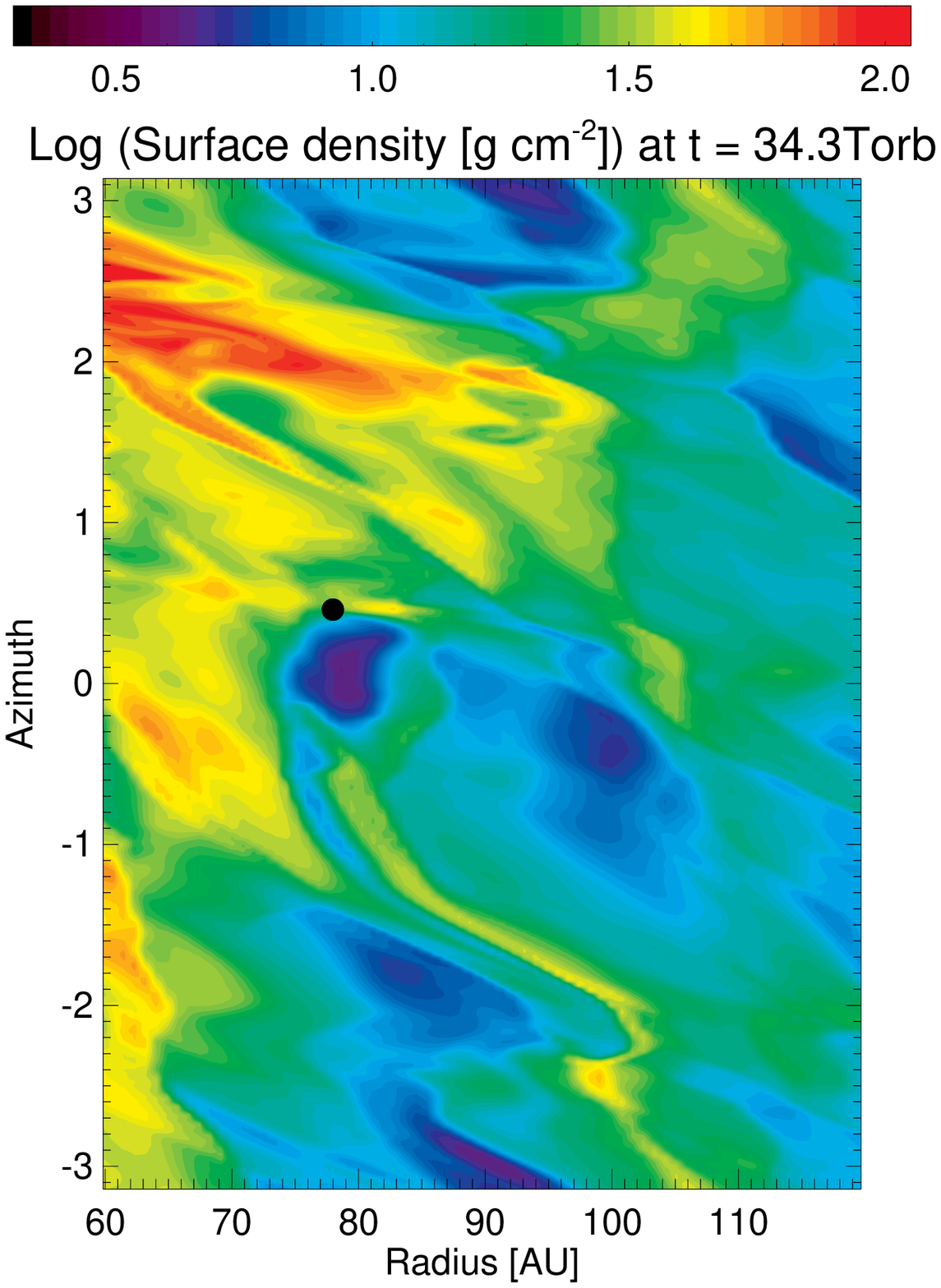}
  \includegraphics[width=0.32\hsize]{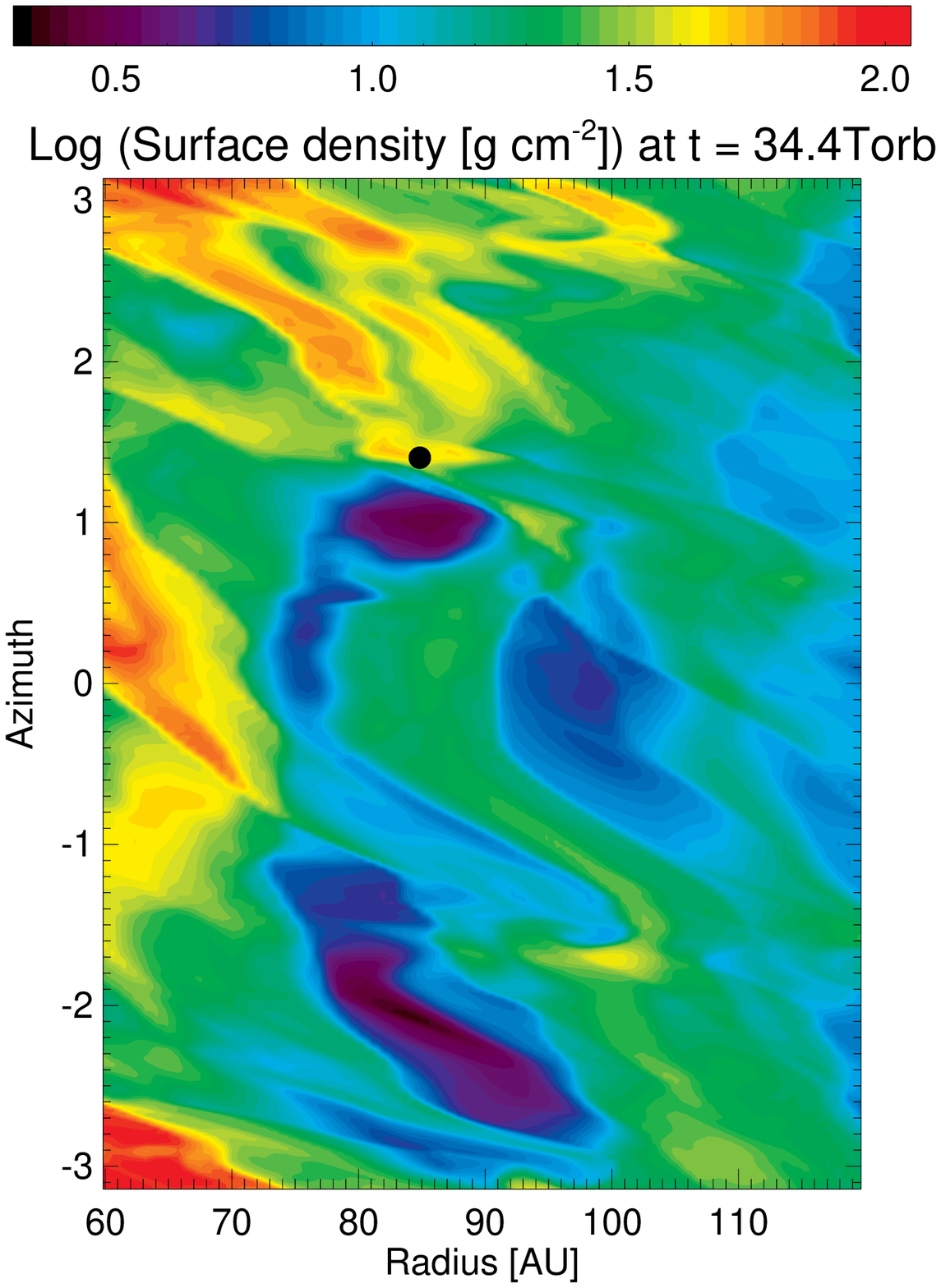}
  \includegraphics[width=0.32\hsize]{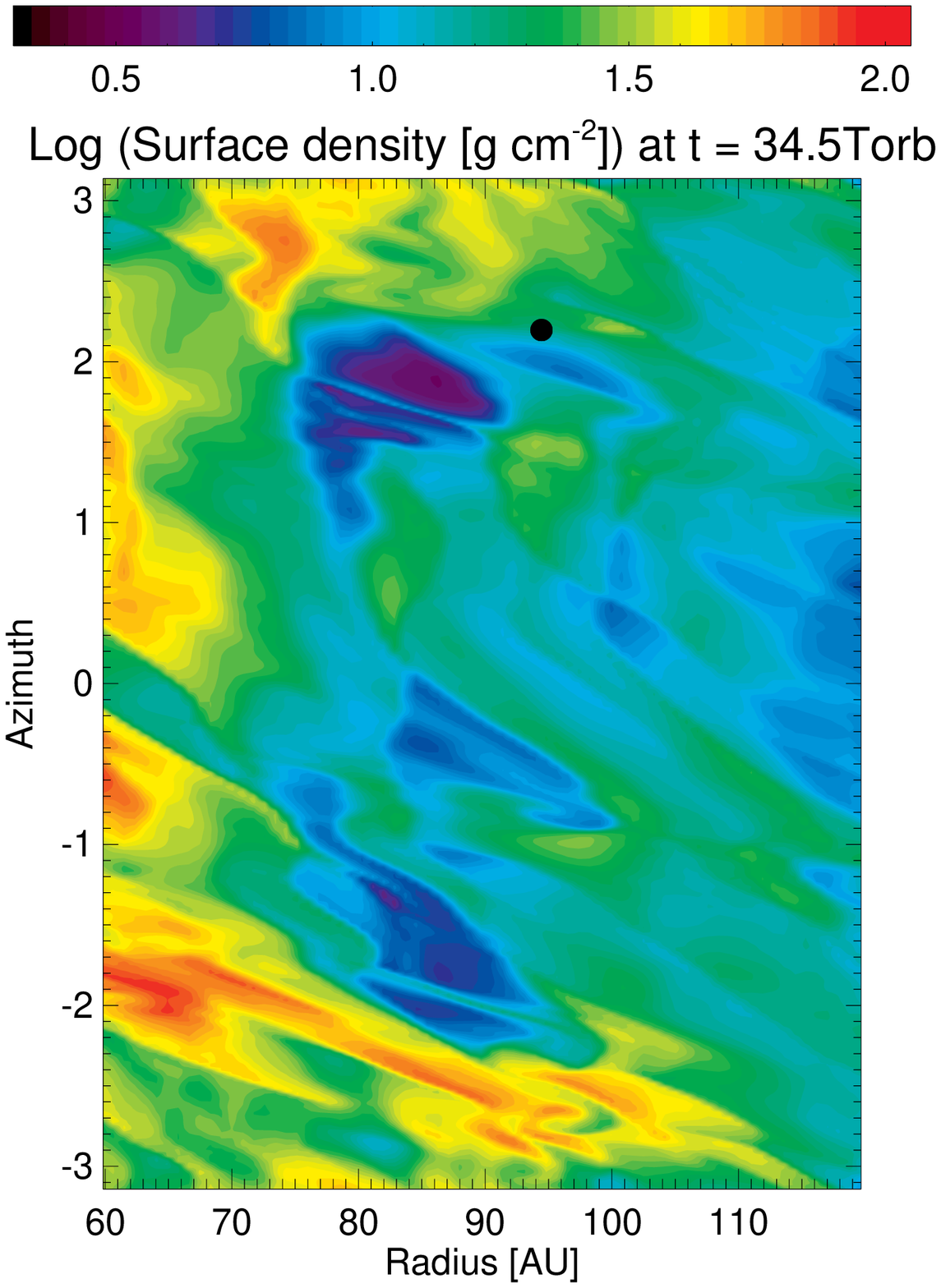}
  \caption{\label{fig:kicks}Contours of the gas surface density
    obtained for the restart simulation with $\beta=30$, and the
    Saturn-mass planet with restart azimuth equal to $7\pi/4$. This
    simulation corresponds to the red curve in the bottom-left panel
    of Fig.~\ref{fig:azimuth}, where a significant episode of outward
    migration occurs from 3 to 6 orbits after restart.  Orbital
    separation is shown on the x-axis, and azimuth on the y-axis.  In
    all panels, the planet's location is depicted with a filled black
    circle.}
\end{figure*}

The net trend coming out of our results is that, in spite of the
stochastic kicks triggered by turbulence, planets migrate inwards very
rapidly. Averaging over $\varphi_p$ and $\beta$, the migration
timescale (defined as the time to migrate from 100 AU to 20 AU, the
location of the grid's inner edge) is typically as short as 3 orbits
for the 5 Jupiter-mass planet, 6 orbits for the Jupiter-mass planet,
and about 8 orbits for the Saturn-mass planet. In addition, at fixed
$\beta$, the amplitude of stochastic kicks increases with decreasing
planet mass, and so does the spread in the migration timescale. Both
results may be qualitatively interpreted as follows. The torque
exerted by the disc on the planet may be decomposed as a background
tidal torque, plus a stochastic torque due to turbulent
fluctuations. In the mass regime that we consider ($q/h_p^3 \sim 1$),
the amplitude of the tidal torque should increase with increasing
planet mass. However, like in discs with turbulence driven by the
magneto-rotational instability, it is unclear if the value of the
tidal torque is similar to that predicted in laminar disc models, in
the absence of turbulence \citep{qmwmhd4}.  A detailed comparison with
results of laminar disc models will be presented in
Section~\ref{sec:laminar}. The standard deviation of the stochastic
torque's distribution is primarily controlled by $\alpha_{\rm tot}$,
and therefore by the cooling parameter $\beta$ (see
Eq.~\ref{alpha_gammie}). For a given $\beta$, the smaller the planet
mass, the smaller the tidal torque to stochastic torque ratio, the
more sensitive the planet is to stochastic kicks, and therefore the
larger the spread in the migration timescale. In the same vein, at
fixed planet mass, the spread in the migration timescale should
increase with decreasing $\beta$. This trend does not clearly show up
in Fig.~\ref{fig:azimuth}, most probably because our minimum and
maximum values of $\beta$ only differ by a factor of 2. Some further
insight into the effect of stochastic kicks is given in
Section~\ref{sec:kicks}. We finally point out that the disc's density
and temperature profiles do not significantly decrease over the
planets migration timescale (for the longest simulation with the
Saturn-mass planet, the disc's mass decreases by at most $10\%$).

\subsubsection{Stochastic kicks}
\label{sec:kicks}
The results of Section~\ref{sec:timescale} show that planets formed by
gravitational instability should rapidly migrate inwards. Even for the
Saturn-mass planet, whose mass is smaller than the estimated initial
clump's mass (see first paragraph of Section~\ref{sec:migration}),
inward migration occurs in typically less than 10 orbits. As
illustrated in Fig.~\ref{fig:azimuth}, planet migration in
gravitoturbulent discs is not a smooth process. Planets may experience
large kicks, either inwards or outwards, depending on the local
density perturbations they encounter. An example of outward kick is
illustrated in Fig.~\ref{fig:kicks}, which displays the gas density
obtained for the restart simulation with $\beta=30$ and the
Saturn-mass planet that experiences a significant episode of outward
migration in the bottom-left panel of Fig.~\ref{fig:azimuth} (red
curve). Time increases by 0.1 orbital periods at 100 AU from left to
right. In each panel, the planet's location is indicated with a filled
black circle.

We first comment that, because the planet's mass is low enough
($q/h_p^3 \approx 0.3$), the wakes it generates are invisible, their
density contrast being much weaker than that of the turbulent
perturbations. In the case depicted in Fig.~\ref{fig:kicks}, the
planet finds itself with an underdense region behind it, and a
(comparatively) overdense region in front of it. This density contrast
yields a large, positive coorbital corotation torque that is
responsible for the vigorous outward kick experienced by the
planet. This kick increases the planet's orbital separation from $\sim
75$ AU to $\sim 95$ AU in only $0.2$ orbital periods at 100 AU (that
is, in 200 years). Similarly, inward kicks are obtained as the planet
experiences a negative coorbital corotation torque induced by an
underdense region ahead of it, and a (comparatively) overdense region
behind it. As is shown for instance in the bottom-left panel of
Fig.~\ref{fig:azimuth}, kicks excite planets epicyclic motion, which
is rapidly damped as the underlying disc--planet tidal torque takes
over.

The stochastic kicks induced by gravitoturbulence are reminiscent of
type III runaway migration. This migration regime relies on the
formation of a coorbital mass deficit in the planet's horseshoe
region, which takes the form of an asymmetric density structure behind
and in front of the planet \citep{mp03}. In massive ($Q \gtrsim 1$)
laminar discs with moderate viscosity ($\alpha$ is a few $\times
10^{-3}$), runaway migration is particularly relevant for planets with
$q \sim h_p^3$ \citep{mp03}, like the planets considered in our
study. However, in our model, the disc's turbulent viscosity
($\alpha_{\rm tot}$ is a few percent) is too large to allow planets to
open a partial dip around their orbit before they reach the disc's
inner edge. Thus, planets do not build up a coorbital mass
deficit. But, through the density fluctuations it triggers,
gravitoturbulence may provide an \emph{effective} mass deficit on each
side of the planet, and induce an effective type III migration.
However, this effective mass deficit is not coorbital with the planet
(as can be seen in Fig.~\ref{fig:kicks}), and therefore cannot lead to
a runaway. Consequently, the relative fast inward migration that we
obtain cannot be attributed to type III runaway migration. Also,
although the planets we consider are massive, their short formation
and migration timescales compared to their gap-opening timescale
suggests they are subject to type I-like migration instead of type II,
perturbed by stochastic kicks. This point will be clarified in
Section~\ref{sec:laminar}.

\begin{figure}
  \includegraphics[width=\hsize]{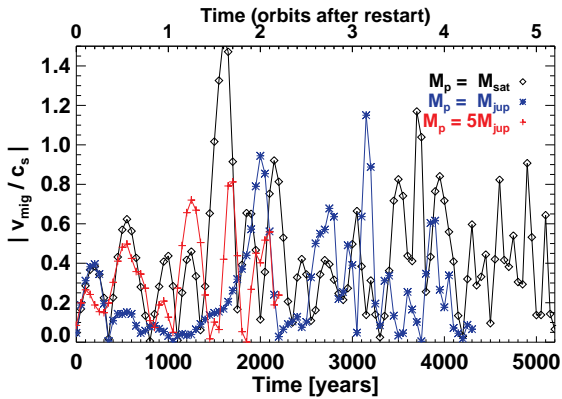}
  \caption{\label{fig:migrate}Time variation of the (absolute value of
    the) migration speed, in units of the local sound speed, obtained
    for the restart simulations with $\beta = 30$ and zero azimuth at
    restart.  They correspond to the results shown with black curves
    in the bottom row of Fig.~\ref{fig:azimuth}. The migration speed
    is measured as the time derivative of the planet's orbital
    separation.}
\end{figure}

Stochastic kicks provide a source of very fast relative motion between
the planet and the disc. Fig.~\ref{fig:migrate} displays the migration
speed in units of the local sound speed for the three runs with
$\beta=30$, and zero azimuth at restart. The migration rate can be a
significant fraction of the local sound speed, and short episodes of
supersonic motion are even obtained. We note that the maximum value
for the migration speed to sound speed ratio tends to increase with
decreasing planet mass, as less massive planets are more sensitive to
stochastic kicks.

\subsubsection{Comparison to laminar disc models}
\label{sec:laminar}
Our results of simulations indicate that the migration of planets
embedded in gravitoturbulent discs is driven by both a vigorous
(negative) tidal torque, and a (positive or negative) stochastic
torque. As the planet mass increases, the former prevails over the
latter. In this section, we compare the results of our turbulent disc
model with $\beta=30$ to those of an equivalent laminar disc model,
including the viscous force in the momentum equation. A constant alpha
viscosity is used, approximately equal to $1.3\%$. The comparison is
done for the same three planet masses as previously considered. As the
initial conditions for the laminar calculations, we take surface
density and temperature profiles that fit those of the
gravitoturbulent run with $\beta=30$ at 30 orbits: $\Sigma \approx
15\;{\rm g\;cm}^{-2}\times (r / 100\;{\rm AU})^{-3/2}$, and $T \approx
25$ K.

Two series of laminar calculations were carried out: (i) one with an
isothermal equation of state, where the imposed (flat) temperature
profile remains constant in time, and (ii) one with an energy equation
with both viscous heating ($\alpha\approx1.3\%$) and $\beta-$cooling
($\beta=30$). For the latter series, the aforementioned\footnote{As in
  the case with $\beta=20$, this value of $\alpha$, given by viscous
  transport theory, is slightly smaller than what we measured in the
  gravitoturbulent disc model ($\langle\alpha_{\rm tot}\rangle \approx
  1.8\%$). We have checked that a higher value of $\alpha$ does not
  change our results of laminar calculations, presented below.}  value
for $\alpha$ leads to an initial thermal balance between viscous
heating and $\beta-$cooling, see Eq.~(\ref{alpha_gammie}).

All laminar calculations include the axisymmetric component of the
disc's self-gravity. This is meant to avoid a large mismatch between
the disc's and the planet's angular velocities due to the disc
gravity, which would yield a spurious shift inwards of all Lindblad
resonances, and therefore an artificial acceleration of the inward
migration \citep{bm08b}. Also, in all laminar runs, the calculation of
the force exerted by the disc on the planet excludes the planet's
circumplanetary disc \citep{cbkm09}. All other parameters are
otherwise the same as in the gravitoturbulent calculations.

\begin{figure}
    \includegraphics[width=\hsize]{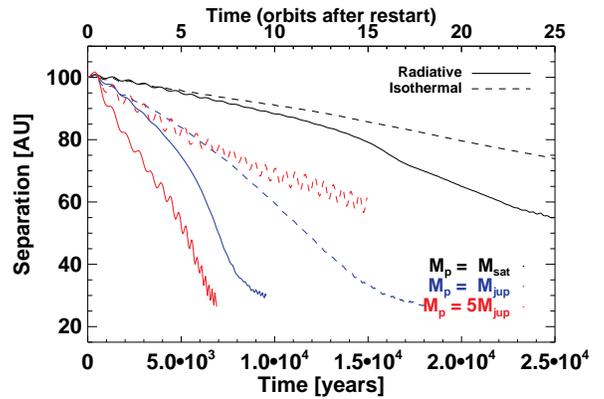}
    \caption{\label{fig:laminar}Time evolution of the planets' orbital
      separation obtained for the laminar disc model with initial
      profiles of surface density and temperature similar to those of
      the gravitoturbulent run with $\beta=30$ before restart.  Solid
      curves show the results with inclusion of an energy equation
      (with viscous heating and $\beta$-cooling), and dashed curves
      those with an isothermal equation of state.}
\end{figure}

The results with the laminar disc model and an energy equation are
depicted in Fig.~\ref{fig:laminar} as solid curves.  It takes about 5
orbits for the 5 Jupiter-mass planet to migrate from 100 AU to 50 AU,
and about 7 orbits for the Jupiter-mass planet. This corresponds to
migration rates $\sim$ half those of the gravitoturbulent calculations
(see bottom row of Fig.~\ref{fig:azimuth}). Interestingly, the
difference in migration rates is larger for the Saturn-mass planet,
which drifts from 100 to 60 AU in $20-25$ orbits in the laminar run,
and (on average) in only $5-6$ orbits in the gravitoturbulent run. We
note that the migration timescales of the Saturn- and Jupiter-mass
planets differ by a factor approximately equal to their mass ratio, as
expected in the type I migration regime, wherein the tidal torque
encompasses the differential Lindblad torque and the horseshoe drag
\citep{pp09a}. We comment that it is the large disc aspect ratio
($h\sim0.1$) that makes the Saturn-mass planet ($q/h^3\sim0.3$) and
the Jupiter-mass planet ($q/h^3\sim1$) relevant to type I
migration. The 5 Jupiter-mass planet ($q/h^3\sim5\gg1$) is, however,
more prone to non-linear effects \citep[e.g.,][]{mak2006}. This is
presumably why the migration timescales of the Jupiter- and the 5
Jupiter-mass planets do not differ by a factor of their mass ratio.
It may seem surprising that the 5 Jupiter-mass planet does not open a
gap and migrates in a very short timescale. The gap-opening criterion
for a planet on a fixed circular orbit in a laminar viscous disc takes
the form \citep{Crida06}
\begin{equation}
  1.1 \left( \frac{q}{h^3} \right)^{-1/3} 
  + 50 \left( \frac{\alpha}{h} \right) \left( \frac{q}{h^3} \right)^{-1} 
  \lesssim 1,
\label{gapcriterion}
\end{equation}
where $\alpha$ and $h$ are to be evaluated at the planet's orbital
separation. Although it is unclear to what extent the gap-opening
criterion in Eq.~(\ref{gapcriterion}) may be applied to planets
migrating in gravitoturbulent discs, it provides some insight into the
fact the planets in our disc models are not expected to clear a
gap. Notwithstanding that the large values of $h$ and $\alpha$ imply
the 5 Jupiter-mass planet does not quite satisfy the above gap-opening
criterion (the left-hand side of Eq.~(\ref{gapcriterion}) typically
ranges from 2 to 4 in the disc inner parts), we stress that the
vigorous tidal torque exerted by the disc, which is partly due to the
large masses of the disc and the planet, conspires to make the planet
migrate in a timescale shorter than the timescale required to open a
gap (a few libration timescales, which would correspond here to a few
tens of orbits). In other words, the 5 Jupiter-mass planet migrates
very fast because it has no time to open a gap.

The results of laminar simulations with an isothermal equation of
state are overplotted as dashed curves in Fig.~\ref{fig:laminar}. The
migration rates obtained in this series are smaller than with
inclusion of an energy equation.  The difference arises from a large,
negative horseshoe drag obtained with an energy equation, and driven
by a \emph{positive} entropy gradient in our disc model \citep[][in
our case, the radial profile of the disc entropy scales in proportion
to the orbital separation]{bm08a, pp08, mc09, pbck10}.

Further insight into the impact of the horseshoe drag may be obtained
by comparing the migration timescales of our laminar calculations to
those predicted by type I migration theory. Note that the predicted
migration timescales are inferred from the torque exerted on a planet
held on a fixed circular orbit. In the isothermal series, where
$\Sigma \propto r^{-3/2}$ and $T$ is uniform, the total torque reduces
to the differential Lindblad torque. Using Eq. (14) of \cite{pbck10},
we find migration timescales of about $75$ and $20$ orbits for the
Saturn- and Jupiter-mass planets, respectively. This is in good
agreement (to within $\sim 25\%$) with our findings (dashed curves in
Fig.~\ref{fig:laminar}). In the series with the energy equation, the
total torque comprises the differential Lindblad torque and the
horseshoe drag. Despite the large value of $\alpha$ in our disc model,
the horseshoe drag is found to be fully unsaturated, as the diffusion
timescale across the planet's horseshoe region remains larger than the
U-turn timescale \citep{bm08a,mc10,pbk11}. Using Eqs. (14) and (45) of
\cite{pbck10}, we now find migration timescales of about $35$ and $10$
orbits for the Saturn- and Jupiter-mass planets, respectively. This is
shorter than what we find (see solid curves in
Fig.~\ref{fig:laminar}), although still in decent agreement. The
difference presumably arises from the large velocity difference
between the disc and the planet, which sets asymmetric horseshoe
streamlines around the planet's orbital radius \citep{masset02}, an
effect that is not taken into account in the horseshoe drag expression
of \cite{pbck10}.

\begin{figure}
  \includegraphics[width=\hsize]{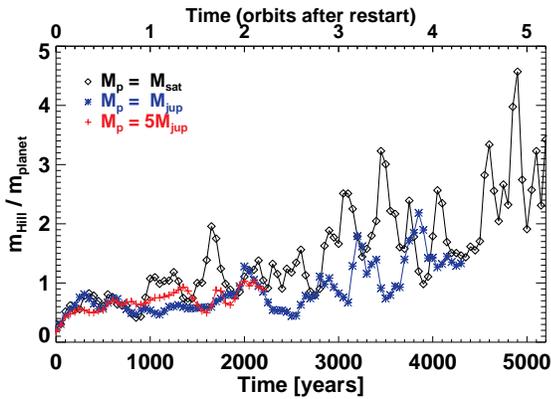}
  \caption{\label{fig:mcpd} Time variation of the mass enclosed in the
    planet's Hill radius in units of the planet mass, obtained for the
    restart simulation with $\beta = 30$ and and zero azimuth at
    restart. The mass ratio is displayed before the planet crosses the
    inner edge of the computational domain.}
\end{figure}
We further comment on having a positive entropy gradient in our
gravitoturbulent disc model. Let us assume the disc profiles of
surface density and temperature scale as $r^{-\sigma}$ and
$r^{-\varpi}$, respectively. The disc entropy $s$, which we define as
$s \equiv p \Sigma^{-\gamma}$, where the disc pressure $p$ verifies
the ideal gas equation of state, scales as $r^{-\xi}$, with $\xi =
\varpi - (\gamma-1)\sigma$. As the disc evolves towards a steady state
with approximately uniform Toomre-Q profile, $\xi \sim -3 +
\sigma(3-\gamma)$. Steady-state gravitoturbulent discs may thus have a
globally decreasing profile of entropy only for surface density
profiles steeper than $r^{-9/4}$, assuming $\gamma=5/3$. Furthermore,
for $\gamma=5/3$ and our fiducial softening length $0.3H(r_p)$, the
fully unsaturated horseshoe drag $\Gamma_{\rm hs}$ reads
\citep[][their Eq. 45]{pbck10}
\begin{equation}
  \gamma \Gamma_{\rm hs} / \Gamma_0 \approx -20.5 + 8.6\sigma,
\label{eqhs}
\end{equation}
where $\Gamma_0 = (q/h_p)^2 \Sigma_p r_p^4 \Omega_p^2$, and all
quantities with the subscript $\emph{p}$ are to be evaluated at the
planet's location. Eq.~(\ref{eqhs}) shows that, in discs with a
uniform Toomre-Q parameter, the horseshoe drag is negative whenever
the surface density profile is shallower than $r^{-2.4}$, assuming
$\gamma=5/3$. The differential Lindblad torque $\Gamma_{\rm L}$ can be
similarly recast as \citep[][their Eq. 14]{pbck10}
\begin{equation}
  \gamma \Gamma_{\rm L} / \Gamma_0 \approx 3.2 - 4.0\sigma,
\end{equation}
and the total torque $\Gamma_{\rm tot}$ as
\begin{equation}
  \gamma \Gamma_{\rm tot} / \Gamma_0 = -17.3 + 4.6\sigma.
\label{eqtotaltq}
\end{equation}
Using Eq.~(\ref{eqtotaltq}), the type I migration timescale, $\tau_I =
r_p^2 \Omega_p M_p / 2 \Gamma_{\rm tot}$, reads
\begin{equation}
  \frac{\tau_I}{T_{\rm orb}} \approx 5.6 \times \left( 3.8 - \sigma\right)^{-1} \gamma Q_p \left( \frac{q}{h_p^3} \right)^{-1} \left( \frac{h_p}{0.1} \right)^{-2},
\label{eq:tauI}
\end{equation}
where $T_{\rm orb}$ stands for the orbital period at the planet's
orbital radius.  The estimate in Eq.~(\ref{eq:tauI}) applies to
planets with $q \sim h_p^3$, and is approximate, since it is based on
the torque expression for a planet on a fixed orbit. As we have
already stressed above, the fast relative motion between the disc and
planet is likely to affect this estimate. Nevertheless, it indicates
that the type I migration of planets formed by fragmentation ($q \sim
h_p^3, Q_p \sim 1.5$) in discs with moderate aspect ratio ($h_p \sim
0.1$) should typically occur in a few tens of orbits at 100 AU, that
is a few $10^4$ years.

\begin{figure*}
\centering
  \includegraphics[width=0.32\hsize]{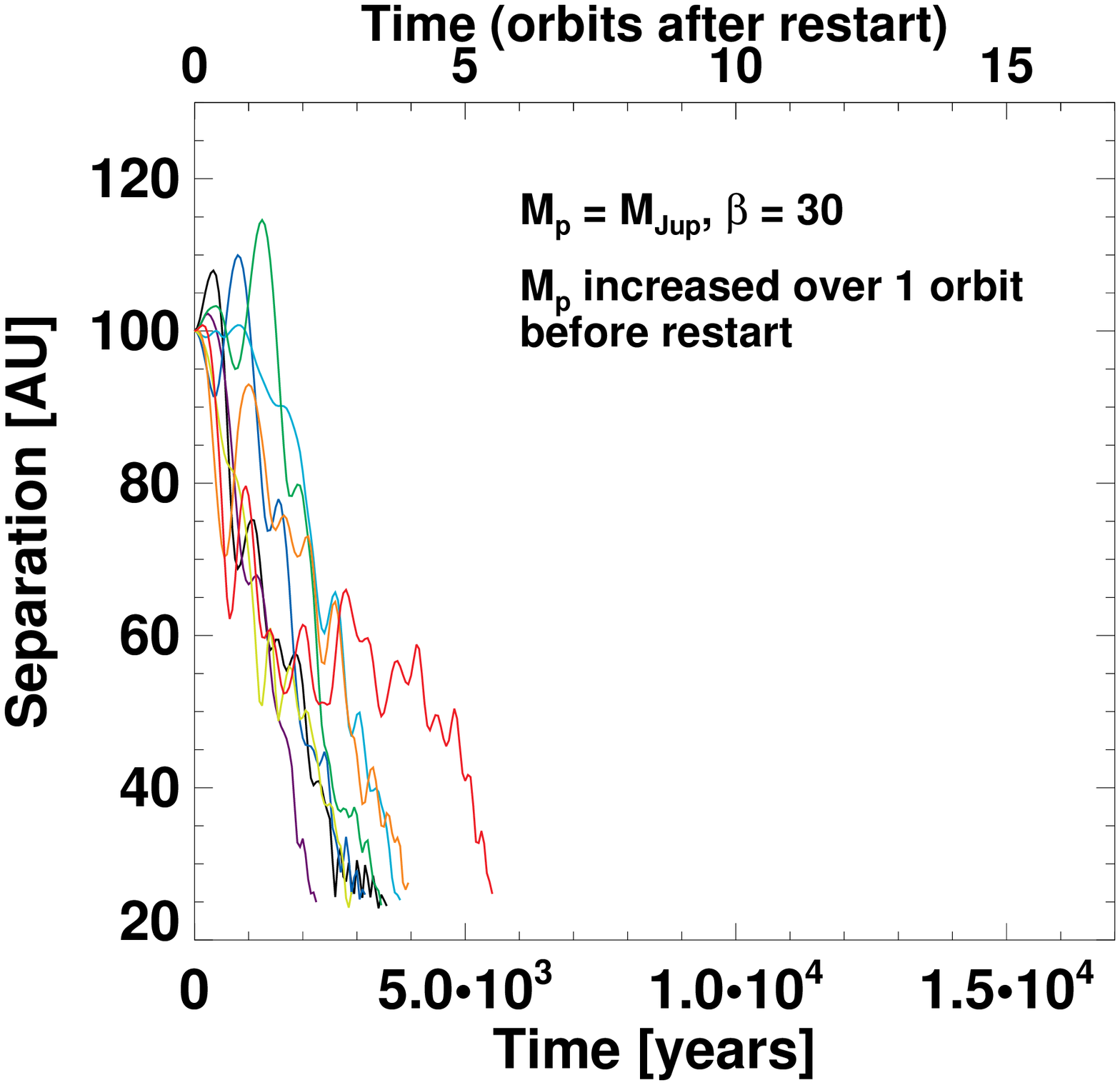}
  \includegraphics[width=0.32\hsize]{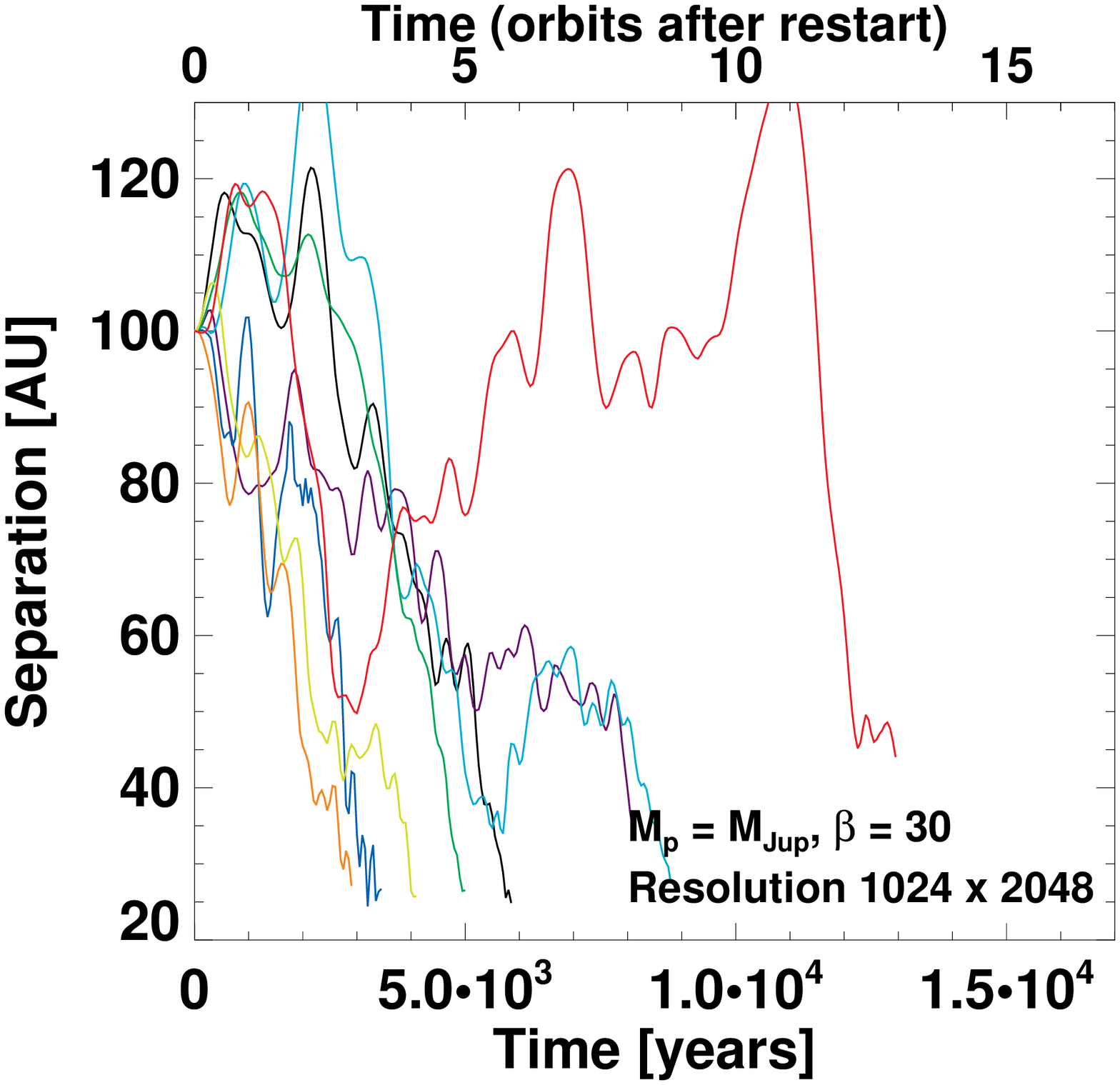}
  \includegraphics[width=0.32\hsize]{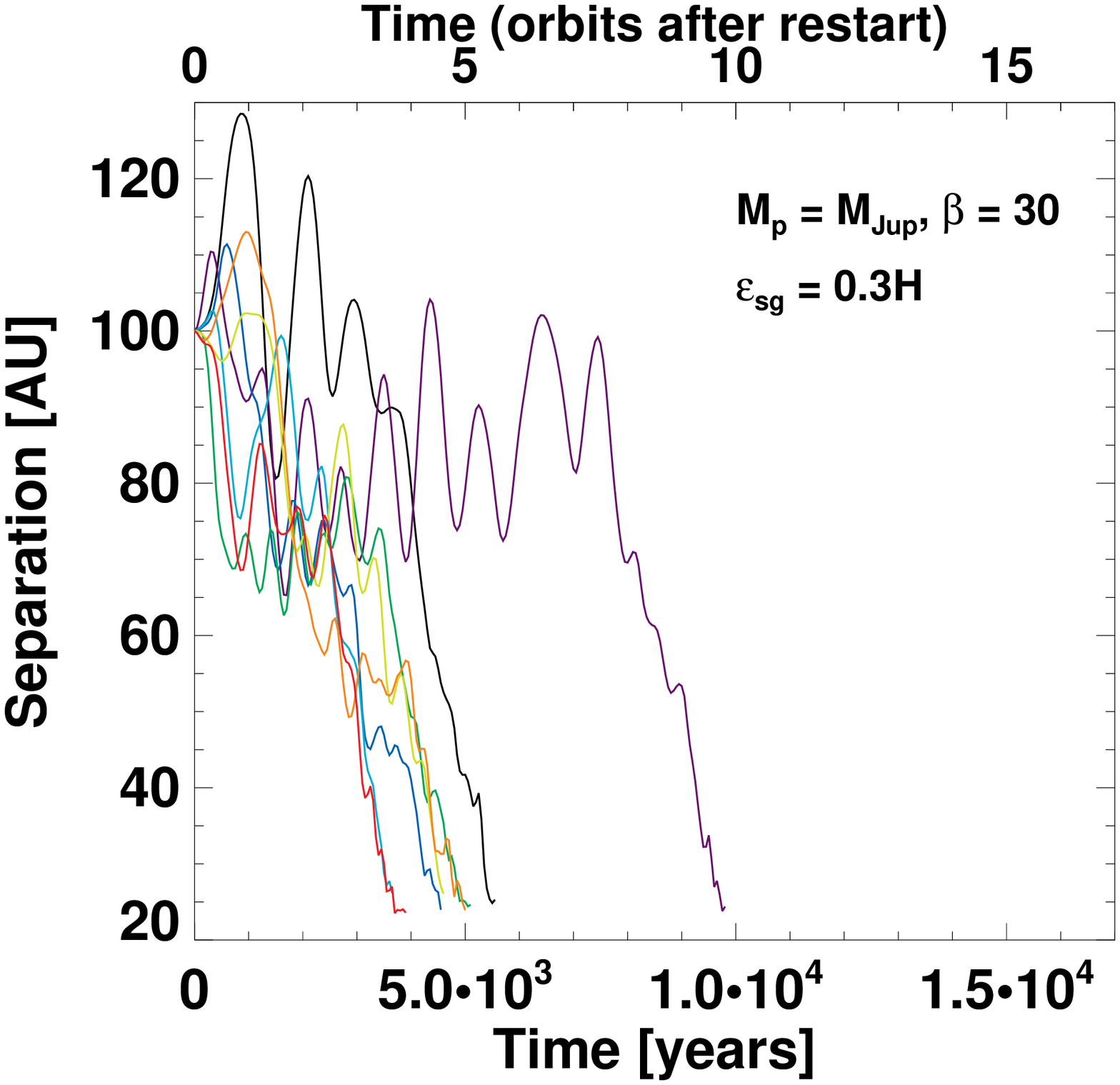}
  \caption{\label{fig:numerics}{Migration of a Jupiter-mass planet in
      a disc with cooling parameter $\beta=30$. In the left panel, the
      planet mass is gradually ramped up over one orbit, while the
      planet is held on a fixed circular orbit. After this first stage
      (not shown in this panel), the planet is released in the disc.
      The middle panel shows results at higher resolution ($1024\times
      2048$), and the right panel shows results with increased
      self-gravitating softening length ($\varepsilon_{\rm sg} =
      3\times 10^{-2}\,r$). All three panels should be compared to the
      mid-bottom panel in Fig.~\ref{fig:azimuth}, which displays the
      results of our fiducial setup (no gradual increase of the planet
      mass, grid resolution is $512\times 1536$, and $\varepsilon_{\rm
        sg} = 3\times 10^{-4}\,r$).}}
\end{figure*}
Still, the migration timescales in our gravitoturbulent simulations
are a factor of a few shorter than obtained in laminar disc models.
In addition to the stochastic kicks due to turbulence, the following
two mechanisms may account for the larger migration rates in the
gravitoturbulent disc model:
\begin{enumerate}
\item[-] The non-axisymmetric component of the disc's self-gravity
  shifts Lindblad resonances towards the planet's orbital radius
  \citep{ph05}. For discs close to marginal stability, and for our
  softening parameter ($\varepsilon / H \sim 0.3$), this shift
  increases the differential Lindblad torque by typically a factor of
  2 \citep[][their Figure 10]{bm08b}. Note however that this mechanism
  has been investigated in laminar disc models. It is therefore not
  straightforward whether it should behave identically with
  turbulence. Still, numerical simulations indicate that, on average,
  the Lindblad torque behaves similarly in laminar and turbulent disc
  models \citep[][where wave-like turbulence is generated by
  stochastic forcing of the disc]{bl10}.
  \medskip
\item[-] In fully self-gravitating discs, the effective mass that
  dictates the strength of the tidal torque and thus the migration
  rate not only includes the planet mass, but also the mass of the gas
  surrounding the planet's location. For massive planets ($q >
  h_p^3$), the latter corresponds to the planet's circumplanetary
  disc, which is a fraction of the planet's Hill sphere
  \citep{cbkm09}. It is unclear what it corresponds to for smaller
  planets; we speculate that it might be the planet's Bondi
  sphere. For illustration purposes, the time evolution of the mass
  inside the planet's Hill radius is shown in Fig.~\ref{fig:mcpd} for
  three gravitoturbulent runs with $\beta=30$. It is comparable to,
  albeit somewhat larger than the planet mass. It progressively grows
  as the planet moves inwards, despite the decrease in the planet's
  Hill radius with decreasing orbital separation. It is possible that
  the short cooling timescales used in our simple cooling prescription
  (a few dynamical timescales, depending on $\beta$) implies that the
  gas in the planet's Hill radius can contract more rapidly than it
  would do with a more realistic cooling function. Again,
  notwithstanding the uncertain dynamical role of the planet's Hill
  sphere in turbulent disc models, it is likely that the increased
  effective mass is responsible for an increase in the migration rate
  by a factor of a few compared to laminar calculations (which only
  include the axisymmetric component of the disc's self-gravity). We
  comment that both the increase in the effective planet mass, and the
  decrease in the disc aspect ratio with radius (in discs with flat
  temperature profile, $h \propto r^{1/2}$), make planets less and
  less sensitive to stochastic kicks as they migrate inwards. This
  explains why in Fig.~\ref{fig:kicks} the amplitude of the stochastic
  kicks decreases with decreasing orbital separation. This also
  implies that, if a planet undergoes a vigorous inward kick, it will
  be increasingly harder to reverse the trend. However, should a
  planet experience a large outward kick, it would take longer for the
  tidal torque to take over and lead to inward migration again.
\end{enumerate}

\subsubsection{Impact of numerics}
\label{sec:numerics}
We briefly assess in this section the impact of numerics on our
results of gravitoturbulent calculations. We have performed an
additional suite of simulations with the Jupiter-mass planet and
$\beta=30$, with the planet's mass first smoothly ramped up over one
orbit, while the planet is held on a Keplerian orbit. After this, the
planet is released in the disc. This procedure is meant to mimic the
clump formation, which typically occurs over a dynamical timescale,
much shorter than the gap-opening timescale (see
Section~\ref{sec:laminar}). The time evolution of the planet's orbital
separation is depicted in the left panel of
Fig.~\ref{fig:numerics}. It shows the very same trend as obtained
without smoothly increasing the planet's mass (see the mid-bottom
panel of Fig.~\ref{fig:azimuth}).

To assess the influence of resolution on our simulation results, we
have varied the grid's resolution, and the softening length
$\varepsilon_{\rm sg}$ over which the self-gravitating acceleration is
smoothed. We have run two additional disc models with $\beta=30$, one
with an increased grid resolution ($1024\times 2048$), and one with a
larger $\varepsilon_{\rm sg}$ ($3\times 10^{-2}\,r$). In the latter
case, $\varepsilon_{\rm sg}$ is 0.3 times the initial pressure scale
height, and it can be regarded as a crude treatment for the effects of
the disc's finite thickness. The value of $\varepsilon_{\rm sg}$ at
$r=r_p$ equals the softening length of the planet's gravitational
potential.  We have checked that before the introduction of the planet
at 30 orbits the disc properties (density and temperature profiles) in
both additional models are in close agreement with those of the
corresponding run at our fiducial resolution and softening length
value. We restarted these simulations with a Jupiter-mass
planet. Again, in each case, a series of eight simulations was
performed with evenly-spaced values for the planet's azimuth at
restart. Results with increased grid resolution, and increased
$\varepsilon_{\rm sg}$ are depicted in the middle and right panels of
Fig.~\ref{fig:numerics}, respectively.  Overall, the amplitude of the
stochastic fluctuations is somewhat larger in the high-resolution
simulations. Still, except in one run that features a rather long
episode of outward migration, the migration timescale of the
high-resolution simulations ranges from about 3 to 9 orbits, which is
broadly consistent with the results at our fiducial
resolution. Furthermore, increasing $\varepsilon_{\rm sg}$ from
$3\times 10^{-4}\,r$ to $3\times 10^{-2}\,r$ does not significantly
change the amplitude of the stochastic kicks experienced by the
planet, and the migration timescales obtained with both values of
$\varepsilon_{\rm sg}$ are in reasonable agreement. The results of
both additional disc models indicate that the planet's orbital
evolution is essentially unaltered by density structures arising from
gravitoturbulence with typical size $\lesssim 0.3H$.
\begin{figure*}
  \centering
  \includegraphics[width=0.32\hsize]{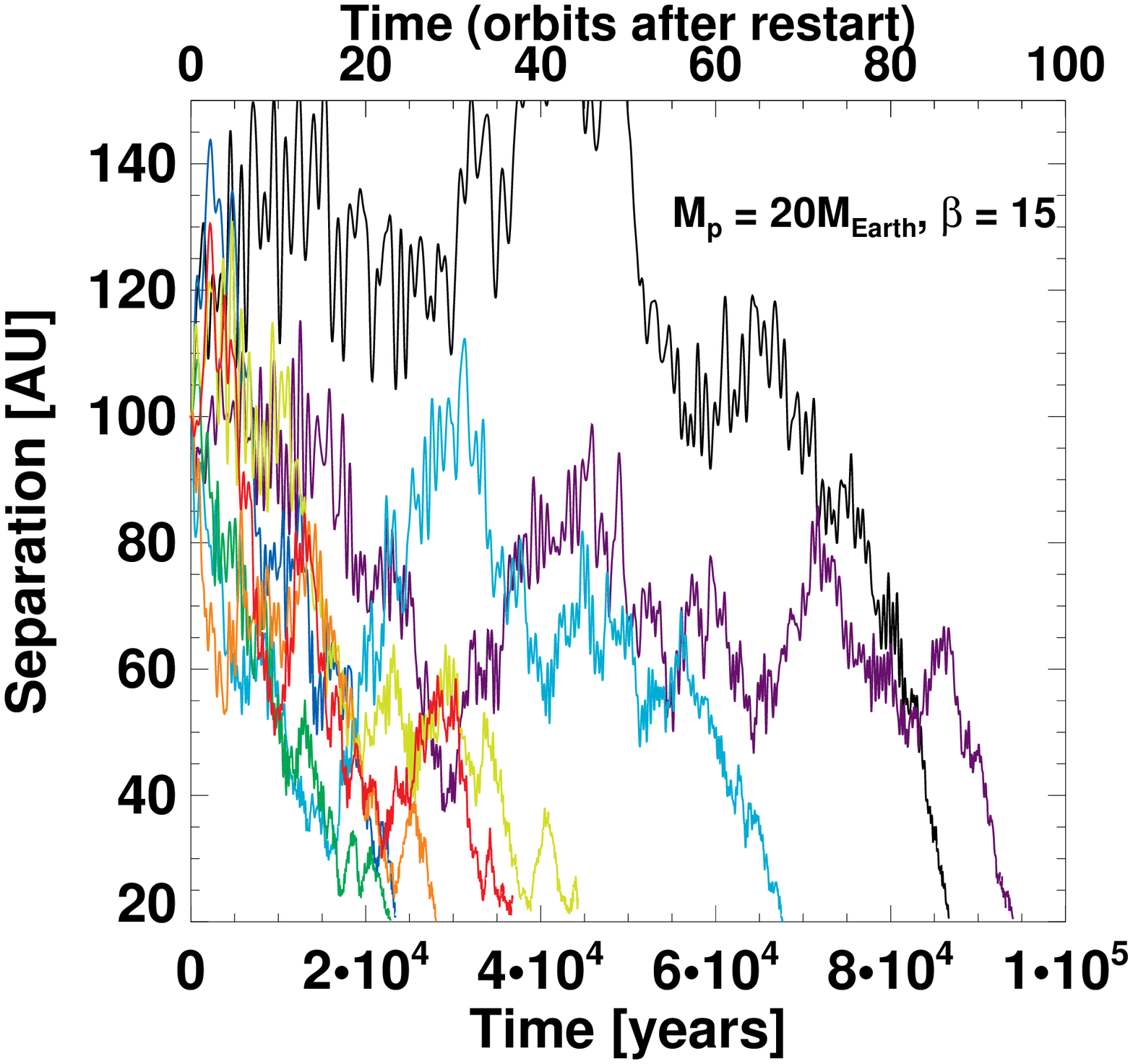}
  \includegraphics[width=0.32\hsize]{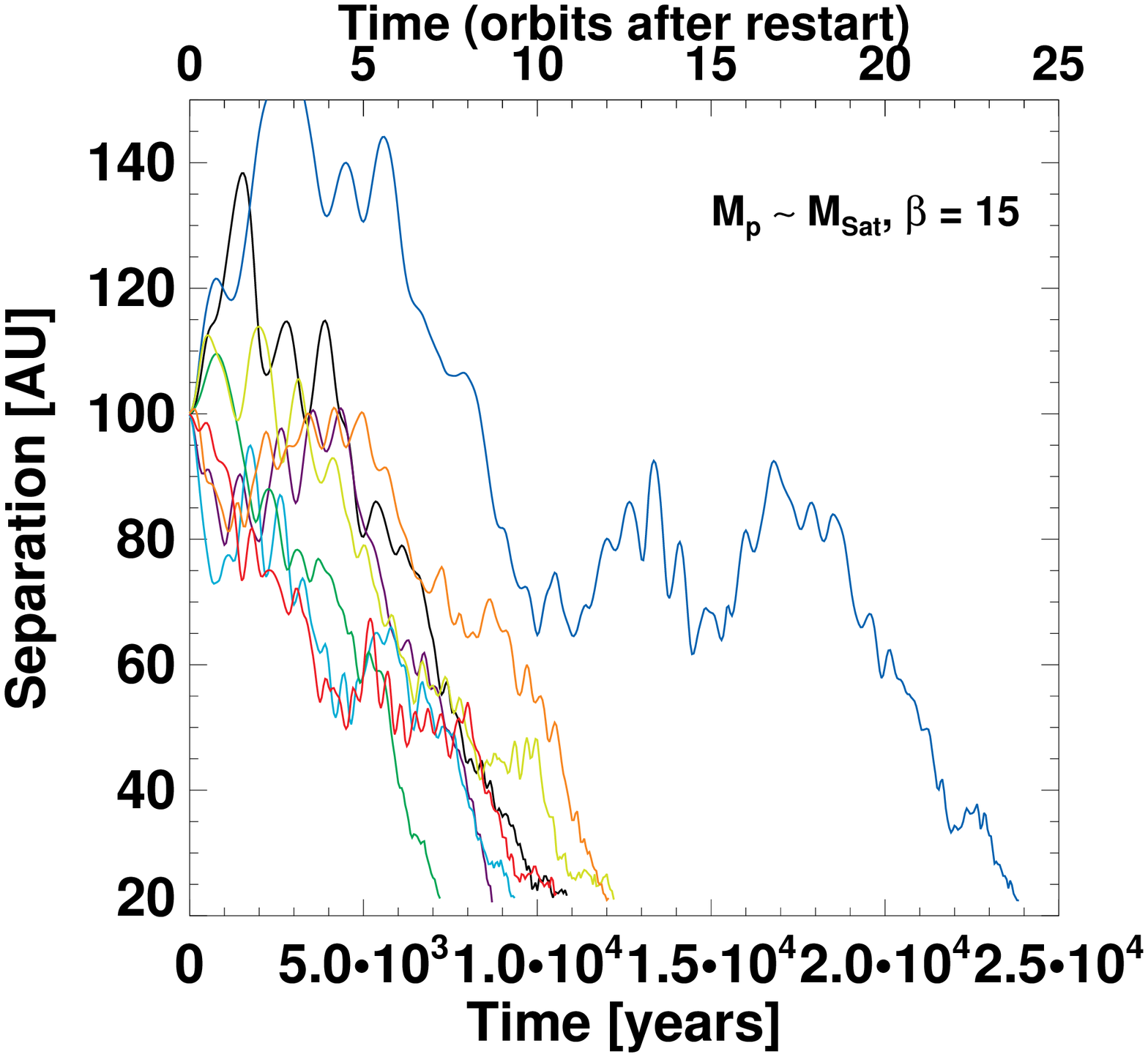}
  \includegraphics[width=0.32\hsize]{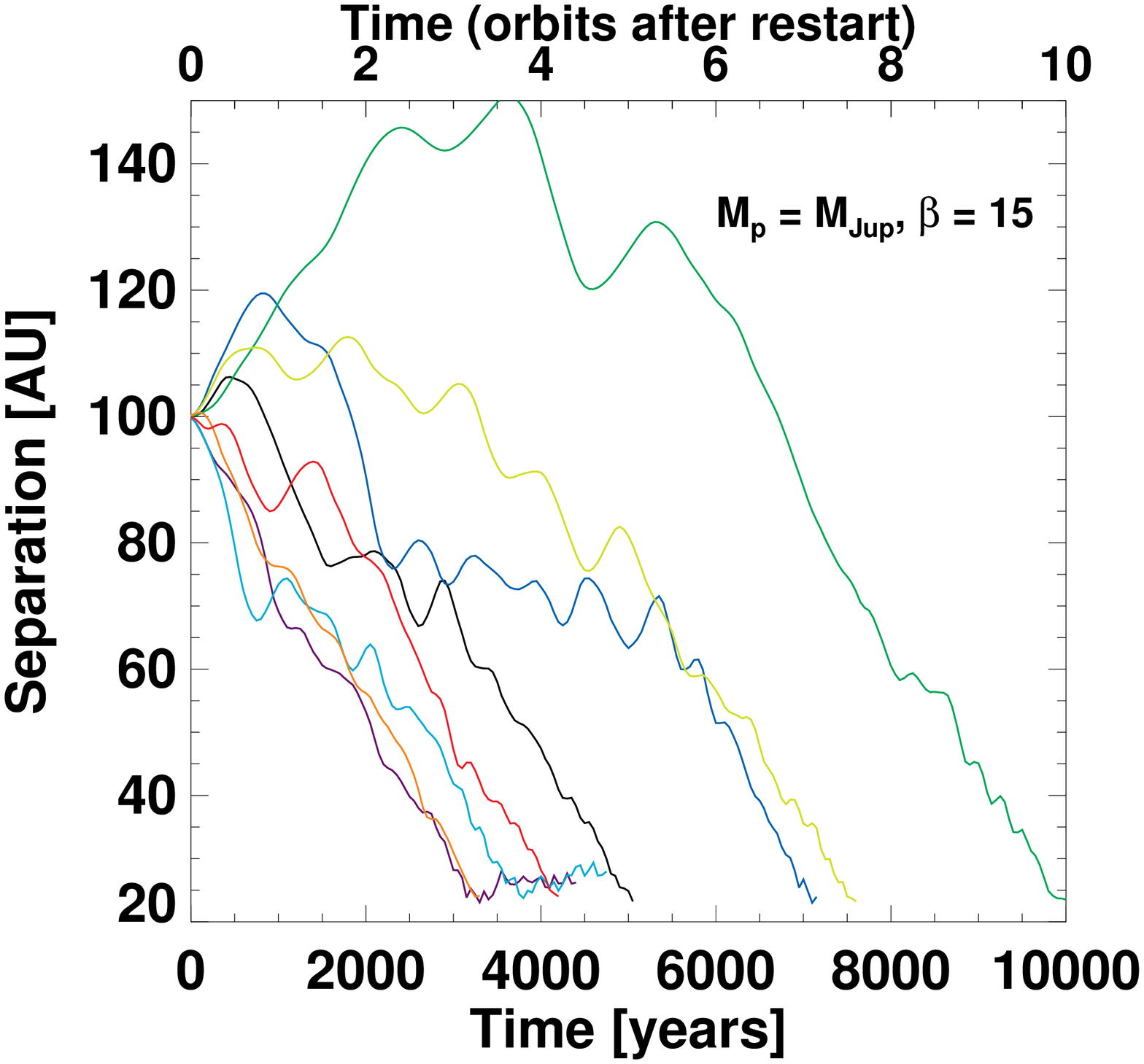}
  \caption{\label{fig:lower}Time evolution of the planets' orbital
    separation for the disc model with $\beta=15$ restarted at 85
    orbits, when the disc aspect ratio is $\approx 6\%$. From left to
    right, results are shown for a 20 Earth-mass planet ($q=6\times
    10^{-5}, q/h^3 \sim 0.3$), a $\sim$ Saturn-mass planet ($q=2\times
    10^{-4}, q/h^3 \sim 1.5$), and a Jupiter-mass planet ($q=10^{-3},
    q/h^3 \sim 5$). As in the previous figures, results are obtained
    for eight evenly-spaced planet's azimuths at restart (increasing
    with increasing wavelength's color).}
\end{figure*}

\subsection{Results at lower disc temperature}
\label{sec:temp}
As shown in Section~\ref{sec:migration}, the migration of planets
formed by gravitational instability results from a competition between
the background disc--planet tidal torque, and the stochastic torque
due to turbulence. At least in the early stages of their evolution,
such planets should have a planet-to-primary mass ratio $q \sim
h_p^3$, and be subject to type I-like migration (see
Section~\ref{sec:laminar}). As far as the tidal torque is concerned,
Eq.~(\ref{eq:tauI}) shows that the migration timescale is essentially
controlled by three dimensionless parameters: $q/h_p^3$, which should
be $\sim 1$ in the early stages of the planet evolution, $Q_p$, which
is $\gtrsim 1.5$ in a steady state, and $h_p$. The disc model that we
have considered so far has $h_p \sim 0.1$, which leads to rapidly
formed Jupiter-mass planets. However, it is possible that the disc
initially has a lower aspect ratio, thereby forming less massive
planets. The latter may still become Jupiter-sized objects if mass
growth is efficient.

To evaluate the impact of a lower disc temperature on our results, we
restarted the simulation of Section~\ref{sec:gts} with $\beta=15$ at
$85$ orbits. At this time, the disc aspect ratio at the planet's
restart location (100 AU) has decreased from 0.1 to about 0.06, and
the disc's mass from $0.2 M_{\star}$ to $0.15 M_{\star}$. We choose
again three planet masses corresponding to a planet-to-primary ratio
$q$ around $h_p^3$: a 20 Earth-mass planet ($q=6\times 10^{-5},
q/h_p^3 \sim 0.3$), a $\sim$ Saturn-mass planet ($q=2\times 10^{-4},
q/h_p^3 \sim 1$), and a Jupiter-mass planet ($q=10^{-3}, q/h_p^3 \sim
5$). As previously, we take eight evenly-spaced planet's azimuths at
restart. The time evolution of the planets' orbital separation is
shown in Fig.~\ref{fig:lower}.

Compared to the results with $h_p = 0.1$, planets with the same
$q/h_p^3$ are now more sensitive to stochastic fluctuations, the tidal
torque becoming less vigorous. This can be seen from
Eq.~(\ref{eq:tauI}), which indicates that at fixed values of $q/h_p^3$
and $Q_p$, the migration timescale due to the tidal torque scales with
$h_p^{-2}$.  The mean migration timescales that we obtain for $h_p
\sim 0.06$ (about 6, 12, and 50 orbits with decreasing $q/h_p^3$) are
typically a factor of 3-5 longer than in the corresponding runs with
$h_p \sim 0.1$, which is roughly consistent with the aforementioned
scaling $\tau_{I} \propto h_p^{-2}$.

\section{Concluding remarks}
\label{sec:conclusion}
Planets observed at large orbital separation from their host star
(typically $\gtrsim 50$ AU) are thought to be potential candidates for
the formation scenario based on the gravitational instability of
massive discs. A number of studies have shown that the formation of
clumps by fragmentation is possible at such separations
\citep[e.g.,][]{Rafikov05, Matzner05, Stamatellos08}. The orbital
evolution of clumps formed by gravitational instability has been
investigated in a few studies \citep[e.g.,][]{Mayer02, boss2005,
  vb10a, cha10, Machida11}. Clumps are often observed to migrate
inwards on short timescales, although in some disc models the
formation of several clumps and/or the use of different cooling
functions has a significant impact on migration, which sometimes
appears to be suppressed \citep{boss2005}.

In this paper, we have studied the interaction between a single planet
and the gravitoturbulent disc it is embedded in, following the
assumption that the planet has formed by gravitational instability. We
have performed 2D hydrodynamical simulations with a simplified
prescription for the disc's cooling. Three disc models with different
cooling timescales have been considered, giving rise to different
levels of gravitoturbulence. After having set up a quasi-steady state
gravitoturbulent disc, we have restarted our simulations including a
single planet with a range of masses approximately equal to the
expected initial mass of clumps formed by fragmentation
\citep[$\gtrsim h^3 M_{\star}$, with $h$ the disc aspect ratio at the
fragmenting location; see][]{Boley10}.

We find that a planet interacting with its nascent gravitoturbulent
disc migrates inwards on very short timescales, despite the stochastic
kicks due to turbulent density fluctuations. In a disc with aspect
ratio $\sim 0.1$, it takes less than 10 orbits at 100 AU (that is,
about $10^4$ years) for a Jupiter-mass planet to migrate from 100 AU
to 20 AU (the inner edge of our computational grid). Planets less
massive than the expected fragmenting mass are more sensitive to
stochastic kicks, but also migrate inwards relatively fast. Their
formation and migration timescales being shorter than their
gap-opening timescale (partly because of the disc's vigorous
turbulence), planets formed by fragmentation do not open a dip or a
gap around their orbit, and are therefore not subject to the type II
and runaway type III migration regimes. Turbulent density fluctuations
may provide an effective mass deficit in the planet's horseshoe region
that kicks the planet either inwards or outwards (see
Section~\ref{sec:kicks}). These kicks can be seen as an effective,
\emph{temporary} type III migration coming on top of the inward
migration due to the background disc--planet tidal torque.

The comparison of our results with those of laminar disc models shows
that the averaged tidal torque driving the net inward migration in
gravitoturbulent discs is essentially the one that would lead to type
I migration in the absence of turbulence (see
Section~\ref{sec:laminar}).  Part of the reason as to why planets
rapidly migrate inwards in our model is due to the set up of a
positive entropy gradient in a steady state, which yields a large
negative horseshoe drag, adding up to the negative differential
Lindblad torque. We argue that in gravitoturbulent discs with uniform
Toomre-Q profiles and cooling-to-orbital timescale ratios (that is,
with a uniform alpha parameter), the horseshoe drag should be
negative, unless the surface density decreases very steeply (see
Eq.~\ref{eqhs}). Although this possibility cannot be excluded, we
believe that fast inward migration should be a generic expectation for
planets formed by gravitational instability.

The main conclusion of this present study is that a single planet
formed by gravitational instability should migrate inwards on a
timescale much shorter than the expected lifetime of protoplanetary
discs. However, it is not possible at this stage to say that massive
planets at large separation from their host star, such as the HR 8799
system, \emph{could not} have formed by gravitational instability. The
simulations presented here contain a number of simplifications that
need to be addressed.  We have considered a simplified cooling
prescription, causing the entire disc to be gravitoturbulent. In
reality, we would expect the inner parts of a disc to be too hot to be
gravitationally unstable, and other sources of turbulence such as the
magnetorotational instability could prevail \citep[see e.g.,][who
studied the radial dependence of the alpha viscosity parameter in
self-gravitating discs with realistic cooling]{Zhu09, RA09,
  Clarke09}. It is thus possible that the rapid type I migration of
planets formed by gravitational instability gets slowed down in the
disc inner parts, resulting in gap formation. A more realistic cooling
prescription will also impact the rate at which planets can contract,
with various outcomes for the evolution of the planet mass over its
orbital migration \citep[e.g.,][]{Nayak10}. Also, we have focused on
the orbital evolution of a single planet assumed to have formed by
fragmentation, whereas the formation of multiple clumps is a likely
outcome of gravitational instability. It is fully expected that the
interaction between clumps may significantly impact their
migration. We will address these issues in future work.

\appendix

\section{Calculation of the alpha parameters in a two-dimensional disc}
\label{sec:alpha}
When mass is confined to a razor-thin disc, the calculation of the
alpha parameters associated with the Reynolds and gravitational
stresses may be inferred from the general three-dimensional case by
substituting the volume mass density $\rho$ with $\Sigma \delta(z)$,
where $\Sigma$ is the surface mass density of the razor-thin disc,
$\delta$ is Dirac's delta function, and $z$ denotes the vertical
coordinate. One obtains \citep{LBK72, Gammie01}
\begin{equation}
  \alpha_{\rm Rey} = \frac{2}{3} \frac{\langle \Sigma \delta v_r \delta v_{\varphi} \rangle}{\langle \Sigma c_{\rm s}^2 \rangle},
\label{alpha_Rey}
\end{equation}
where $\langle . \rangle$ stands for the azimuthal average, $\delta
v_{r} = v_{r} - \langle v_{r} \rangle$, $\delta
v_{\varphi} = v_{\varphi} - \langle v_{\varphi} \rangle$, and
\begin{equation}
  \alpha_{\rm sg} = \frac{2}{3} \frac{\langle \int_{-\infty}^{\infty} (4\pi G)^{-1} g_r g_{\varphi} dz \rangle}{\langle \Sigma c_{\rm s}^2 \rangle}.
\label{alpha_sg}
\end{equation}
In both previous equations, $v_r$, $v_{\varphi}$ and $c_{\rm s}$ are
to be evaluated in the razor-thin disc ($z \equiv 0$) and are
therefore directly computed in a 2D simulation. However, the vertical
dependence of $g_r$ and $g_{\varphi}$, the radial and azimuthal
components of the self-gravitating acceleration, needs to be
determined as the gravitational field outside the disc contributes to
the shear stress despite mass being confined to the disc. The
potential $\Phi$ from which $g_r$ and $g_{\varphi}$ can be derived is
given by
\begin{equation}
  \Phi = -\int_{\cal{D}} \frac{G\Sigma(r^{\prime} ,\varphi^{\prime})r^{\prime}dr^{\prime}d\varphi^{\prime}}{\sqrt{r^2 + r^{\prime 2} - 2rr^{\prime} \cos(\varphi-\varphi^{\prime}) + z^2 + \varepsilon^2_{\rm sg}}}
\end{equation}
where $\cal{D}$ is the domain where $\Sigma$ does not vanish, and
$\varepsilon_{\rm sg}$ is a small softening length required to avoid
numerical divergences when $z=0$. From the surface density of our 2D
simulations, $g_r(r,\varphi,z)$ and $g_{\varphi}(r,\varphi,z)$ can be
conveniently calculated by fast Fourier transforms provided that $z$
is taken proportional to $r$, condition required for $g_r$ and
$g_{\varphi}$ to read as convolution products (another necessary
condition is that a grid with logarithmic radial spacing is used). The
same condition applies to $\varepsilon_{\rm sg}$ for the calculation
of $g_r$ and $g_{\varphi}$ at $z=0$. Writing $\varepsilon_{\rm sg} = B
r$ and $z = \eta r$, with $B$ and $\eta$ constants, the expressions
for $g_r(r,\varphi,z)$ and $g_{\varphi}(r,\varphi,z)$ are given by
Eqs.~(A1) and~(A3) of \cite{bm08b} with $B^2 \rightarrow B^2 +
\eta^2$. The vertical integral in Eq.~(\ref{alpha_sg}) is then
numerically calculated from the values of $g_r$ and $g_{\varphi}$
obtained with a range of values for $\eta$.

\section*{Acknowledgments}
CB is supported by a Herchel Smith Postdoctoral Fellowship.  FM is
grateful for funding from the EC Sixth Framework Programme and also
acknowledges the support of the German Research Foundation (DFG)
through grant KL 650/8-2 within the Collaborative Research Group FOR
759: {\it The formation of Planets: The Critical First Growth Phase}.
SJP is supported by an STFC Postdoctoral Fellowship. Numerical
simulations were performed on the Pleiades and Grape clusters at
U.C. Santa Cruz. Part of this project was done during the 2010 edition
of the ISIMA summer school, whose support is gratefully
acknowledged. We thank John Papaloizou and Andrew Youdin for useful
discussions, Min-Kai Lin for detailed comments on a first draft of
this manuscript, and the referee for useful comments.

\bibliographystyle{mn2e}

\label{lastpage}

\end{document}